%% file: main.tex
\documentclass{layout/epfl-report}

\usepackage{csquotes}       

\usepackage{biblatex}
\setlist{itemsep=-2pt} 

\begin{document}

\include{macros}

\frontmatter

\title{Synthetic\\Photography\\Detection}
\subtitle{A Visual Guidance for Identifying Synthetic Images Created by AI}
\author{M. Mathys, M. Willi, R. Meier}

\affiliation{University of Applied Sciences and Arts Northwestern Switzerland, armasuisse S+T} 
\coverimage{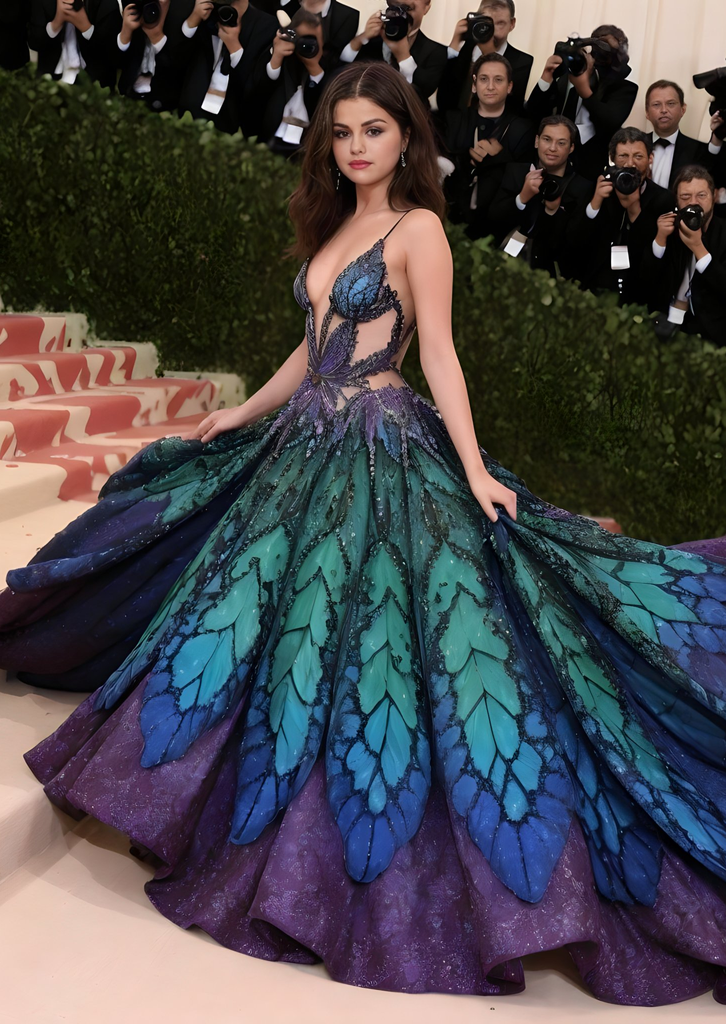}
\definecolor{title}{HTML}{FFFFFF} 

\makecover

\tableofcontents


\mainmatter

\chapter{Introduction}
\label{chapter:introduction}

Artificial Intelligence (AI) tools have become incredibly powerful and are beginning to shape many aspects of daily life. These tools can be used to find information quickly \cite{reid_generative_2024}, converse with virtual assistants \cite{mickle_apple_2024}, enhance and manipulate smartphone photos \cite{berrada_4_2023}, or recommend personalized playlists on music streaming services \cite{jockims_how_2024}. Significant advancements in generating synthetic photographs that appear convincingly real, e.g. \cite{podell_sdxl_2023}, are one of the most recent developments. \\

Photorealistic synthetic images can be used in various contexts, including advertisements, creative works, or simply for entertainment. However, alongside such applications, more sinister uses have emerged, such as employing synthetic images for deceptive purposes. This deception can occur in commercial settings and may range from innocuous (e.g. creating and selling generated content \cite{kelly_how_2023}) to malicious (e.g. selling non-existing/fake products \cite{tenbarge_etsy_2024}). On a grander scale, synthetic images may be used to influence the opinions and views of individuals, groups, or even entire societies \cite{karnouskos_artificial_2020}. It is thus crucial to empower consumers of digital media to recognize synthetic photographs to avoid harm and misinformation.\\

\input{figures/intro_example}

Although AI tools have become increasingly adept at generating synthetic photographs of remarkable quality, they are not without flaws. While these images may seem convincing at first glance, many inconsistencies become apparent upon closer inspection. For example, Figure \ref{fig:intro_example} shows a photograph purportedly depicting a scene from the Israel-Hamas war. Despite of its unknown origin, it was widely shared across social media platforms. While the intention behind creating and disseminating this image is unclear, it is evidently of synthetic origin. As highlighted in the figure, there are numerous inconsistencies or, as we refer to them in this article, \textit{artifacts}. Artifacts are telltale signs that an image is not authentic (i.e. fully or partially synthetic).\\

We have catalogued common flaws and present various examples that illustrate failures of generative AI. We consider synthetic images from the state-of-the-art family of diffusion models, such as \cite{rombach_high-resolution_2022, podell_sdxl_2023, noauthor_midjourney_nodate}. By highlighting image artifacts and presenting the range of defects that synthetic images may exhibit, we hope to direct viewers towards elements that are particularly susceptible to revealing their synthetic nature and to equip them with the knowledge and tools necessary to discern between real and synthetic images. \\

We define a taxonomy of artifact categories (Chapter \ref{chapter:artifact_types}), provide various examples of artifacts that might appear in synthetic images (Chapter \ref{chapter:spotting_the_artifacts}), discuss challenges when assessing images for artifacts (Chapter \ref{chapter:challenges}), and provide suggestions for practical applications and future research (Chapter \ref{chatper:what_now}). Most of the synthetic images, if not indicated otherwise, were created by ourselves (see Appendix for details) and depict scenes that might be used in the context of deceptive purposes.

\chapter{Artifact Types}
\label{chapter:artifact_types}

Artifacts are \enquote{defects in an image (such as a digital photograph) that appear as a result of the technology and methods used to create and process the image}\footnote{\href{https://www.merriam-webster.com/dictionary/artifact}{Merriam-Webster dictionary}}. They play a crucial role in traditional image forensics and in identifying images of synthetic origin \cite{cardenuto_age_2023, guillaro_trufor_2023}.\\ 

Artifacts can be divided into different categories, such as \textit{physical} and \textit{geometrical} inconsistencies, or problems in the representation of \textit{human anatomy} and \textit{text}, or inexplicable \textit{distortions} in the image, as well as implausible \textit{semantic} scene composition. Figure \ref{fig:artifact_overview} shows one example for each category.

\input{figures/artifacts_one_of_each}

Figure \ref{fig:artifact_taxonomy} shows a more detailed overview where some of the categories have been further subdivided. We were inspired by the works of \cite{borji_qualitative_2024, cao_synartifact_2024} to define this taxonomy. The purpose of this taxonomy is to provide a tool for the systematic analysis and identification of artifacts in potentially generated images. We note that artifacts might reasonably be assigned to multiple categories and that the categories themselves might overlap to some degree in how they are defined. \\

\textbf{Physical} artifacts comprise all aspects or elements in an image that contradict or violate the laws of physics, especially those of optics and gravity. Models learn to generate physically plausible images indirectly by being exposed to billions of images, which inevitably leads to inaccuracies. Particularly challenging is optics: the accurate depiction of reflections \cite{lin_detecting_2024, cardenuto_age_2023}, light sources and how they affect the depicted scene \cite{farid_lighting_2022}, object shadows and their consistency across a scene \cite{borji_qualitative_2024, lin_detecting_2024, cao_synartifact_2024, sarkar_shadows_2023}, gravity-related issues which can manifest in objects floating without physical support \cite{borji_qualitative_2024,lin_detecting_2024}, as well as implausible behavior of fluids, cloth or other materials \cite{borji_qualitative_2024}. \\

\textbf{Geometrical} artifacts refer to incorrect representations of objects concerning their shape, size, surface structure, and perspective. These issues arise when generative models fail to accurately depict geometrical properties, resulting in distorted or inconsistent shapes, such as desks with wavy surfaces or uneven leg lengths and numbers (see examples in \cite{borji_qualitative_2024}). Scale and perspective \cite{farid_perspective_2022} errors also occur, where objects are not shown in proper spatial relationships to each other, leading to unrealistic scenes where closer objects may appear smaller than distant ones. Depth-related inconsistencies, such as improper layering of objects and inaccurate texture gradients, further diminish realism. Symmetry issues can cause naturally symmetrical objects to appear asymmetrical, and repetitive patterns may be inconsistently rendered, resulting in misaligned bricks or irregular window patterns. These artifacts typically stem from limitations in the models' training data or architectural design, affecting their ability to apply principles of projective geometry and maintain consistency in textures and patterns.\\

\textbf{Human Anatomy} includes complex structures that pose significant challenges for generative models, which, as a result, struggle to accurately recreate them. These artifacts are particularly noticeable and disturbing due to humans' innate ability to detect deviations in human forms. Common issues include unnatural asymmetries in faces, such as irregularities in eyes \cite{borji_qualitative_2024, lin_detecting_2024, porcile_finding_2023}, teeth, hair, and facial expressions. Hands often exhibit problems like duplicated or merged fingers, incorrect numbers of fingers, and unrealistic poses, highlighting a well-known difficulty in generating realistic hand images \cite{delaney_why_2023}. The overall body and limbs may appear with unnatural proportions or poses, and skin tones can seem inconsistent or artificial. Accessories and wearables, such as glasses and jewelry, often display asymmetries, missing parts, or unnatural integration with the body, further diminishing the realism of synthetic images. \\

\textbf{Distortions} are categorized as noisy or inconsistent areas that are inherent to the generated image, rather than characteristics of specific objects. These include color-related distortions, such as inconsistent colors, unusual color choices, and improper saturation. Noise and pattern interference can manifest as semi-regular noise like banding or checkerboard patterns, and areas may appear blurred or smeared. Blurring and detail loss are common, especially in backgrounds, making objects and individuals unrecognizable and creating inconsistencies in detail levels. Stylistic anomalies may also occur, where images intended to be photorealistic instead resemble cartoons or paintings. \\

\textbf{Text} artifacts are common and manifest in incomprehensible, duplicated, or scrambled depictions of text, as well as spelling errors \cite{borji_qualitative_2024,cao_synartifact_2024,cardenuto_age_2023}.
More subtle forms of text artifacts might manifest in inconsistent typefaces, i.e. different instances of the same letter in the same text may differ in size or appearance.\\

\textbf{Semantic} artifacts do not violate physical laws or defy conventional expectations regarding an object's geometry, yet still fail to convey realism. This can be due to issues related to the technical functionality, intended use, identity of a person, or cultural or social context. They can be characterized as errors in constructing a plausible scene or inaccuracies in depicting the true relationships between objects \cite{borji_qualitative_2024}.

\input{figures/artifacts_overview}

\chapter{Spotting the Artifacts}
\label{chapter:spotting_the_artifacts}

In the following we are going to examine synthetic images of different types and from different scenes. These images might exhibit different artifact types. Depending on the image type, some categories might be more relevant than others.

\section{People}

Accurate depiction of people, particularly, human anatomy, is difficult. Such images might be used in social media profiles to impersonate someone or to create fake identities \cite{yang_characteristics_2024}. \\

\textbf{Faces, Heads, and Portraits} might exhibit specific artifacts related to human anatomy. Figure \ref{fig:people_hair} shows a synthetic photography which mimics a professional headshot. This example is rather convincing on first glance. However, closer inspection reveals subtle artifacts related to eye shape, hair strands, jewelry (wearables), clothing, and teeth detail, however, barely perceptible.

\input{figures/people_intro}

\textbf{Eyes} seem challenging to accurately generate by deep learning models. Figure \ref{fig:people:eyes} depicts examples of synthetic photographs with a spotlight on the eyes. These generated eyes may exhibit a range of anomalies such as inconsistent reflections, unnatural shapes, or asymmetries that deviate from the norm. However, it is important to note that natural eyes can also present a variety of irregularities. Real human eyes are not always perfectly symmetrical and may have slight differences in shape, size, and orientation. Additionally, eyes can show unique characteristics such as minor asymmetries, varied pupil shapes, and differing levels of light reflection. These natural variations can sometimes make it difficult to distinguish between real and synthetic images based solely on the appearance of the eyes. 

\input{figures/people_eyes}

\textbf{Teeth} may also provide clues if a photograph is synthetic. Figure \ref{fig:people_teeth} shows synthetic teeth which can lack clear boundaries between individual teeth. In generated images, teeth often appear fused together or blurred, lacking the distinct separations and details that characterize natural teeth. These synthetic representations may exhibit a uniform, overly smooth appearance that fails to capture the subtle variations and textures found in real teeth. Natural teeth, on the other hand, exhibit clear boundaries between each tooth, with noticeable gaps and contours that define their individual shapes. While natural teeth are generally distinct, they can also be irregular, displaying a range of characteristics such as varying sizes, slight misalignments, and unique shapes. These irregularities contribute to the authenticity and realism of a photograph. The examples provided in Figure \ref{fig:people_teeth} highlight the differences between synthetic and natural teeth, illustrating how the lack of precise detailing and individual separation in generated images can serve as an indicator of their artificial nature. \\

\input{figures/people_teeth}

\textbf{Ears and wearables} may be rendered with unnatural asymmetries and shapes. Figure \ref{fig:people_ears_wearables} depicts examples where ears are deformed or asymmetric, although this can be difficult to judge without careful scrutiny. In generated images, ears often show irregularities such as unusual shapes, inconsistent sizes, and positioning that deviates from natural human anatomy. These deformations can manifest as uneven lobes, misplaced angles, or other distortions that disrupt the natural symmetry typically observed in real human ears.

Additionally, wearables such as earrings and glasses can also exhibit telltale signs of synthetic generation. Asymmetrical earrings, for instance, may appear mismatched or improperly aligned, with one earring significantly different in size, shape, or position compared to the other. Glasses, too, may not sit naturally on the face, showing irregularities in the alignment of the frames or inconsistencies in the way they rest on the ears and nose.

These anomalies are particularly important to note because wearables are expected to be symmetrical and properly fitted in real-life scenarios. The examples provided in Figure \ref{fig:people_ears_wearables} highlight these issues, showing how generated images can fail to accurately replicate the symmetry and proper fit of accessories like earrings and glasses.

\input{figures/people_ears}

\textbf{Hair} in generated images often presents anomalies that distinguish it from naturally photographed hair. Common issues include unnatural textures, local blur, incomplete hair strands, asymmetries in hairstyle, and unnatural placement and behavior. Unnatural textures can make hair appear too smooth, shiny, or uniformly unrealistic. Local blur may cause parts of the hair to seem out of focus or smeared, particularly around the edges or where it interacts with other elements like the face or clothing. Incomplete hair strands can appear as clumps or blocks of color, lacking the fine details of individual strands. Asymmetries in hairstyle may show odd or inconsistent patterns, such as uneven lengths or irregular partings. Unnatural placement and behavior can make hair appear to float or defy gravity and fail to respond appropriately to the light source in the image. Figure \ref{fig:people_hair} illustrates these various artifacts, showing how even advanced models struggle to replicate the complexity of human hair.

\input{figures/people_hair}

\textbf{Groups of People} are often more challenging for generative models to accurately depict, as compared to close-up photographs of individuals. This is due to additional anatomical details which need to be rendered, such as hands, arms, legs, and feet. Additionally, there is more room for error related to the scene composition and semantic inconsistencies. We have also observed that people in any given image often look similar, sometimes even alike. Figure \ref{fig:people_group_individual} shows detailed specific artifacts related to hair, hands and eyes. \\

\input{figures/people_groupe_single}

Figure \ref{fig:people_refugees_group} also shows multiple synthetic images where visible artifacts can easily be spotted, such as malformed hands.\\

The images in Figures \ref{fig:people_group_individual} and \ref{fig:people_refugees_group} take up the theme of the image in Figure \ref{fig:intro_example}, showing desperate refugee families with small children. Such images can easily be used maliciously to garner attention, elicit emotions, and manipulate opinions---all the more so if they are mistaken for real photographs.\\

\input{figures/people_group_war}

\textbf{People in the background} often lack detail, particularly their faces. Figure \ref{fig:people_meta_gala_gomez} shows a synthetic photograph of Selena Gomez supposedly participating at the Met Gala 2024. While Selena Gomez is depicted relatively convincingly, the photographers in the background indicate undoubtedly that this image is not real. The faces of these background figures lack essential details, appearing blurry or indistinct. Additionally, their facial expressions often appear strange or inconsistent with what one would expect in a natural setting. These anomalies can manifest as unnatural smiles, misplaced features, or an absence of the subtle variations and nuances that characterize real human expressions. Additionally, anatomical irregularities occur often, such as disfigured hands or body parts merging with other objects.

\input{figures/people_met_gala_gomez}

Moreover, the background people’s attire and body language may also show irregularities. Poses may seem stiff or awkward, further hinting at the synthetic nature of the image. The cameras held by the photographers in Figure \ref{fig:people_meta_gala_gomez} are another indicator of the image’s synthetic origin. These devices may exhibit geometric artifacts, such as distorted shapes, improper alignments, unnatural proportions that do not match real-world cameras, or are merged with the photographer's hands.

\clearpage

\section{Indoor Scenes}

Indoor scenes often exhibit artifacts related to lighting, reflection, geometry, and semantic inconsistency. Synthetic images of apartment rooms could for example be misused in fake advertisements to deceive potential tenants or buyers. As an example, Figure \ref{fig:indoors_example} shows a synthetic photograph of a bathroom with a myriad of artifacts: an irregular water tap that appears misshapen, a drawer with only one visible edge, a disfigured shower head with an unnatural shape, and mirror images that do not match with the actual objects in the room.\\

\input{figures/indoors_example}

Images of interiors typically feature numerous right angles and straight, parallel lines. Real photographs of three-dimensional spaces represent a 3D-to-2D projection, where parallel lines on the same plane all converge to the same vanishing point \cite{farid_perspective_2022}. Figure \ref{fig:indoor_vanishing_points_2} illustrates this effect: While the lines of the real photography intersect at the same point, the lines of the synthetic version fail to do so, which constitutes an artifact of perspective geometry.\\

\input{figures/indoors_vanishing_point_2}

Semantic inconsistencies are another indicator of synthetic images. Figure \ref{fig:indoor_apartments} depicts photographs of apartment rooms that exhibit numerous semantic and geometric artifacts: irregularly shaped windows and wall sections that defy typical construction norms, misshaped kitchen utensils that look unnatural and out of place, inconsistent perspective that distorts the spatial arrangement, and irregular floor tiling with patterns that do not align properly.\\

\input{figures/indoors_apartments}

Lighting artifacts are also prevalent in synthetic images of indoor scenes. Figure \ref{fig:indoor_bunkers} depicts synthetic images of underground bunkers with artificial light sources. In these images, the light sources often do not behave as expected: Shadows may fall in incorrect directions, light intensity might be unevenly distributed, and the overall lighting can appear artificial. These inconsistencies in lighting can disrupt the realism of the image and provide clues that the scene is not genuine. However, identifying inconsistencies in lighting and shadows may be challenging for laypeople (including the authors of this article). \\

The images of underground bunker rooms shown in Figure \ref{fig:indoor_bunkers} are inspired by the various synthetic images related to the Israel-Hamas conflict that have recently been circulating on social media, such as Figure \ref{fig:intro_example} or a synthetic image of a supposed underground storage of weapons next to donated medical equipment\footnote{\href{https://x.com/reddit_lies/status/1729698746961707348}{X/Twitter}}.
\\

\input{figures/indoors_underground}

\clearpage
\section{Outdoor Scenes}

Outdoor scenes challenge generative models in different aspects, particularly if objects are present that are usually regular in shape and have a degree of symmetry. Figure \ref{fig:outdoor_intro} shows an example of an image, allegedly showing an explosion or fire near the Pentagon. It was shared widely on social media \cite{clayton_fake_2023} and supposedly caused a short drop in the stock markets. In addition, the Arlington Fire \& EMS Department felt compelled to publish a statement saying that no such incident was taking place\footnote{\href{https://x.com/ArlingtonVaFD/status/1660653619954294786}{X/Twitter}}. The image is relatively easy to identify as synthetic as it contains many obvious flaws related to the geometry, particularly non-uniformity of objects. Building facades are irregular, as well as the fence along the street. Furthermore, the street lamp post is partly in front (as it should be) and partly behind the street fence. There are also garbled textures and distortions present. The image was shared in different quality levels, low quality versions are more difficult to recognize as synthetic.The version shown in Figure \ref{fig:outdoor_intro} is the highest quality version that we were able to find and allows for zooming in to clearly visible artifacts.\\

\input{figures/outdoor_intro}

Issues with perspective and vanishing points are also prevalent in images from outdoor scenes. Figure \ref{fig:outdoor_vanishing_points} illustrates a scenery which is particularly susceptible to such artifacts: streets and building facades with numerous parallel lines. These lines should converge to the same vanishing point, maintaining consistent perspective geometry. However, in synthetic images, these parallel lines often deviate, failing to meet at a common point, which disrupts the natural appearance of the scene. This deviation is a clear indicator of synthetic origin. Careful inspection of the perspective in outdoor scenes, especially those with many straight, parallel lines, can reveal these subtle artifacts.\\

\input{figures/outdoor_vanishing_point}

Figure \ref{fig:outdoor_war_zone_city} depicts war scenes in the street of a city with destruction, fires and people running to safety (a theme similar to Figure \ref{fig:outdoor_intro}). The model\footnote{\href{https://civitai.com/models/178910/sdxl-cctv}{sdxl-cctv}} that was used mimics images from cctv which explains the higher camera angle in the images. Different artifacts related to scale, detail and shape symmetry can be observed.  \\

\input{figures/outdoor_war_zone}

\clearpage
Figure \ref{fig:outdoor_water_park} shows synthetic images from fictional places that were shared widely on social media \cite{riebeling_santorini_2024}. The images exhibit many artifacts upon closer inspection, such as incorrect human anatomy, blurred faces, as well as illogical geometrical shapes. Figure \ref{fig:outdoor_fake_vacation} also depicts fictional photographs of vacation places and hotels that contain geometrical issues as well as physical artifacts in the form of incorrect reflections.\\

\input{figures/outdoor_water_park}

\input{figures/outdoor_fake_vacation}

\clearpage
\section{Objects}
\label{sec:spotting:objects}

Synthetic photographs of objects might be difficult or easy to identify as such, depending on expected symmetries, regularities, and the domain knowledge of the analyst.\\

Objects like vehicles, which are often designed with strict symmetry along various axes, provide ideal opportunities to identify synthetic images. This is because perfect symmetry is challenging for generative models to replicate, as it is a feature learned from data rather than inherently encoded in the architecture of these models. Consequently, even small deviations in symmetry can serve as strong indicators of a synthetic origin.\\

Figure \ref{fig:objects_himars} displays a synthetic depiction of a High Mobility Artillery Rocket System (HIMARS)\footnote{\href{https://en.wikipedia.org/wiki/M142_HIMARS}{Wikipedia: M142 HIMARS}}. Such an image could be used to create and disseminate synthetic military vehicle images to mislead opponents in armed conflicts. The asymmetries present in this image offer robust indications that it is synthetic. For example, the irregular tire profile is highly unusual for such a vehicle. Additionally, there are noticeable left/right asymmetries in the shape of the reflectors and the lights, which are typically identical in a real HIMARS. These inconsistencies highlight the challenges that generative models face in creating perfectly symmetrical objects.\\

\input{figures/objects_himars}

Figure \ref{fig:objects_vehicles} shows synthetic images of different vehicles. The bicycle is a particularly difficult object to generate, as the spokes of the wheels must be highly regular and the gear that transmits the motion from the pedals to the rear wheel must appear technically functional. In the depicted synthetic image, however, the spokes are irregular and uneven, deviating from the expected precision, and the gear seems disfigured and appears blurry. Furthermore, vehicles that may not be well-represented in the training data of generative models, such as helicopters, can exhibit unrealistic geometrical forms. These deviations from realism, such as misshapen rotor blades or inconsistent body proportions, serve as additional clues that the image is synthetic.\\

\input{figures/objects_vehicles}

Domain knowledge plays a crucial role in identifying synthetic images of objects. Experts familiar with the typical features and specifications of certain objects can more easily spot anomalies that might be overlooked by the untrained eye. For instance, an automotive engineer might quickly notice discrepancies in the design of a car's bodywork, such as irregularities in the symmetry of the panels, the alignment of the wheels, or the placement of lights and mirrors. These professionals are trained to recognize the precise and often standardized elements that make up a vehicle, and any deviation from these standards can indicate a synthetic image.
Similarly, an aviation expert could identify unnatural shapes or configurations in synthetic images of aircraft. Such experts might look for inconsistencies in the proportions of wings and fuselage, the positioning and size of windows, or the alignment of the landing gear. The deep understanding of the design and engineering principles that govern these vehicles enables these professionals to detect subtle inaccuracies that generative models might produce. \\

Figure \ref{fig:objects_items} illustrates various items, some of which appear highly realistic. For example, the crochet puppy dog in the figure is by nature irregular and therefore difficult to identify as synthetic. The handcrafted appearance with its inherent imperfections makes it challenging to distinguish from a real object. In such cases, domain knowledge related to textile crafts and handmade items might aid in the assessment, helping to spot inconsistencies in the stitching patterns or the texture of the yarn. The depiction of the PC tower in Figure \ref{fig:objects_items} offers semantic clues regarding its synthetic origin. An expert in computer hardware might quickly notice the absence of essential components, such as a missing power supply unit, irregular port configurations, or unrealistic placement of internal components. These kinds of semantic errors are often overlooked by generative models but stand out to those with specific knowledge in the domain. \\

\input{figures/objects_items}

In summary, while synthetic images can be highly convincing at first glance, the presence of anomalies and inconsistencies becomes apparent when viewed through the lens of domain-specific expertise. This underscores the importance of combining technical tools with expert human judgment in the ongoing effort to identify and mitigate the spread of synthetic imagery.

\chapter{Challenges}
\label{chapter:challenges}

Many of the synthetic images presented are seemingly easy to identify because they exhibit clear artifacts that are readily visible and detectable upon close inspection. However, identifying synthetic images is not always straightforward due to various challenges.

\section{Difficult Examples}

The most recent generative models can produce synthetic photographs with very subtle or ambiguous artifacts, making correct detection challenging and prone to error. Difficult examples often closely mimic the characteristics of authentic photographs. Figure \ref{fig:people_face_difficult} is an example of a particularly high quality synthetic image of a face. \\

\input{figures/people_face_difficult}

Figure \ref{fig:difficult_items} shows images from a class of synthetic photographs that are often difficult to correctly classify: hand crafted objects in which small irregularities appear perfectly natural and can even enhance the impression of realism. \\

\input{figures/difficult_items}

To correctly detect synthetic photographs, more sophisticated analysis techniques may be used to augment the visual inspection. Image analysis tools \cite{guillaro_trufor_2023} or AI algorithms \cite{wang_cnn-generated_2020} can help analyze images for invisible artifacts. As generative models continue to improve, we expect more ambiguous cases, no matter what analysis tools and techniques might be applied.\\

Performing a thorough context analysis can also help analyze difficult examples. Understanding the circumstances under which the image was taken, its source, and its intended use can provide critical context. Cross-referencing the image with other available data and performing consistency checks can further aid in determining its authenticity. Leveraging domain-specific knowledge, as discussed in previous chapters (e.g. Section \ref{sec:spotting:objects}), becomes crucial. Experts familiar with the subject matter depicted in the image can provide valuable insights and identify anomalies that may not be apparent to the untrained eye.

\section{Mind the Bias}

When analyzing an image, one should be concerned with two possible mistakes: false positives and false negatives. A false positive occurs when an authentic image is incorrectly identified as
synthetic. This type of error can be caused by unwarranted skepticism and may discredit legitimate photographs, causing misinformation and mistrust. On the other hand, a false negative occurs when a synthetic image is mistakenly identified as authentic. This error can have serious implications, especially in contexts where the authenticity of the image is crucial, such as in news reporting, legal evidence, or scientific research. To minimize the risk of such errors, it is important to be aware of potential biases  \cite{sunde_cognitive_2019, rudiger_cognitive_2016} that may influence our analysis and to avoid jumping to hasty or overconfident conclusions.\\

One of the cognitive biases that can significantly affect image analysis, is confirmation bias \cite{forensic_science_regulator_forensic_2020, casu_ai_2023}. Confirmation bias is the tendency to search for, interpret, and remember information in a way that confirms one’s preconceptions or hypotheses. In the context of identifying synthetic images, confirmation bias can lead an analyst to overlook genuine artifacts in authentic images or to see artifacts where none exist in synthetic images. For example, if an analyst expects an image to be synthetic, they might scrutinize it more closely and interpret any minor irregularities as evidence of its synthetic nature, even if those irregularities are within the range of natural variations. Conversely, if an analyst believes an image to be authentic, they might dismiss or rationalize visible artifacts, attributing them to photographic errors or imperfections.\\

Background and expert knowledge can also increase bias \cite{dror_linear_2021}, such as unfounded prejudices, expectations, or overconfidence \cite{somoray_providing_2023}. For this reason, background knowledge should only be revealed and used at a later stage of the assessment process, after an unprejudiced analysis of the image has already been carried out \cite{dror_linear_2021}. \\

Being aware of cognitive biases, such as confirmation bias, and employing a systematic approach to image analysis can help mitigate the risk of false positives and false negatives. Possible mitigation techniques include educating analysts about the existence and potential effects of cognitive biases to increase their awareness, initially observing \enquote{what's there} without expecting or searching for specific details, and consistently looking for arguments that both support and contradict any assumptions or conclusions \cite{forensic_science_regulator_forensic_2020, sunde_cognitive_2019}. By combining advanced detection tools, domain-specific knowledge, and contextual analysis, we can improve our ability to accurately distinguish between authentic and synthetic images.

\chapter{What Now?}
\label{chatper:what_now}

Our goal is to provide guidance on how to identify synthetic imagery to a wide variety of interested persons, such as forensic and intelligence analysts or simply consumers of digital media. We aim at increasing the resilience of individuals, groups, and society with respect to malicious attempts to deceive, defraud, and manipulate them with synthetic material. To this end, the contents of the previous chapters might be used to create teaching material or to device strategies regarding the analysis of suspicious photographs.

\section{Practical Applications}
\medskip
\textbf{Education and Training}: Educators can use the compiled material to develop training programs for fields such as digital forensics, intelligence analysis, journalism, and media literacy. Workshops and courses can be designed to teach participants in how to recognize and analyze artifacts in synthetic images, while mitigating any potential biases. By simulating real-world scenarios, these training sessions could help individuals practice and refine their detection skills.\\

\textbf{Forensic \& Intelligence Analysis}: Forensic and intelligence analysts can apply the taxonomy and examples provided in this article to improve their methods for investigating photographs. By incorporating the artifact categories into their analysis framework, they can systematically assess images for signs of manipulation or synthesis. This structured approach carries the potential to enhance the accuracy and reliability of their findings.\\

\textbf{Media Literacy Campaigns}: Public awareness campaigns can utilize the information from this article to educate the general public about the prevalence and risks of synthetic images as well as their current shortcomings in form of the various artifacts presented. In addition, campaigns should also address the challenges arising from a potentially increased distrust among the general public towards all digital media (also called imposter bias \cite{casu_ai_2023}). This may come from the increasingly widespread availability and ease of use of generative AI to synthesize media content. By promoting critical thinking, reflective media use, and skepticism, these campaigns could increase societal resilience against malicious uses of synthetic images.\\

\textbf{Policy Development}: Policymakers can leverage the insights from this research to highlight the complexity of distinguishing real from synthetic photographs, illustrate challenges, and ultimately motivate the development of regulations and guidelines aimed at curbing the misuse of synthetic images. By establishing standards for digital content verification and encouraging transparency in image creation, they can help mitigate the impact of synthetic images on society.

\section{Future Research Directions}
\medskip
\textbf{Advancing Detection Technology}: While it is crucial to empower individuals with the means to distinguish real from synthetic photographs, it is clear that manual image verification does not scale to a large amount of images. Hence, we consider detecting synthetic images with technology, such as AI tools, a crucial area of research \cite{wang_cnn-generated_2020, epstein_online_2023}. Furthermore, visible artifacts might become harder to identify by visual inspection as generative models improve, increasing the relative importance of detection technology.\\

\textbf{Understanding Cognitive Bias}: Further studies on cognitive biases in image analysis can provide deeper insights into how these biases affect decision-making. By developing strategies to counteract these biases, researchers can improve the reliability of image authentication processes.

\setcounter{biburlnumpenalty}{7000}
\setcounter{biburllcpenalty}{7000}
\setcounter{biburlucpenalty}{7000}

\printbibliography[heading=bibintoc] %


\appendix

\chapter{Sources of Synthetic Images}
\label{appendix:image_sources}

Many of the synthetic images were generated by ourselves. The following model checkpoints were used:

\begin{itemize}
    \item\href{https://civitai.com/models/133005}{Juggernaut XL}
    \item \href{https://civitai.com/models/178910/sdxl-cctv}{SDXL CCTV (LoRA)}
    \item \href{https://civitai.com/models/25694/epicrealism}{epiCRealism}
    \item \href{https://civitai.com/models/4201/realistic-vision-v60-b1}{Realistic Vision V6.0 B1}
    
\end{itemize}

As indicated in figure captions, some images are from the Synthbuster dataset \cite{bammey_synthbuster_2023} which is licensed under \href{https://creativecommons.org/licenses/by-sa/4.0/deed.en}{CC BY-SA 4.0}.

\end{document}

%% file: macros.tex

\newcommand{\overviewminipagewidth}{0.25\textwidth}

\newcommand{\bboxcolor}{cyan}
\newcommand{\bboxbordersize}{thin}
\newcommand{\bboxcornerformat}{rounded corners}

\newcommand{\makerectanglecorners}[5]{
  \path let \p1=(#1.north east) in coordinate (firstcorner) at (\x1*#2, \y1*#3);
  \path let \p2=(#1.north east) in coordinate (secondcorner) at (\x2*#4, \y2*#5);
}

\newcommand{\makebboxline}[2]{
\draw[\bboxcolor, \bboxbordersize] (#1) -- (#2);
}

\newcommand{\makethicklinewithcolor}[3]{
\draw[#3, line width=2mm] (#1) -- (#2);
}

\newcommand{\makethickdottedlinewithcolor}[3]{
\draw[#3, dotted, line width=2mm] (#1) -- (#2);
}

\newcommand{\makethindottedlinewithcolor}[3]{
\draw[#3, dotted, line width=0.3mm] (#1) -- (#2);
}

\newcommand{\makearrow}[5]{
  \path let \p1=(#1.north east) in coordinate  (firstpoint) at (\x1*#2, \y1*#3);
  \path let \p2=(#1.north east) in coordinate (secondpoint) at (\x2*#4, \y2*#5);
  \draw[->, line width=5mm, \bboxcolor, \bboxbordersize, anchor=#1.south west] (firstpoint) -- (secondpoint);
}

\newcommand{\drawbbox}[4]{
\node[inner sep=0, draw=\bboxcolor, \bboxbordersize] (bbox) at (#1) {
\begin{tikzpicture}[inner sep=0]
\clip(#1) rectangle (#2);
\node[inner sep=0] {\includegraphics[width=#4]{#3}};
\end{tikzpicture}
};
}

\newcommand{\makeandplacecrop}[6]{
    \path let \p3=(#1.north east) in coordinate (position) at (\x3*#2, \y3*#3);
    \node[inner sep=0, scale=#6, draw=\bboxcolor] (cropbox) at (position) {
        \begin{tikzpicture}[inner sep=0]
            \clip(firstcorner) rectangle (secondcorner);
          \node[inner sep=0] {\includegraphics[width=#5]{#4}};
        \end{tikzpicture}
        };
}

\newcommand{\drawbboxlabel}[6]{
\path let \p1=(#1.north east) in coordinate  (upperleftcorner) at (\x1*#2, \y1*#3);
\path let \p2=(#1.north east) in coordinate  (contactpoint) at (\x2*#4, \y2*#5);
\node[draw=\bboxcolor,  \bboxbordersize, \bboxcornerformat, text=black, text opacity = 1.0, fill=white, fill opacity=0.5, align=left] (textbox) at (upperleftcorner) {#6};
\draw [\bboxcolor, \bboxbordersize] (textbox) -- (contactpoint);
}

\newcommand{\imagewithcropbelow}[7]{%
    \begin{subfigure}[b]{#6}
        \center
        #7
        \begin{tikzpicture}[anchor=south west]
            \def\imagewidth{\textwidth}
            \def\imagesource{#1}
            \def\imagename{imagelinkedin}
            \node[name=\imagename, inner sep=0]{\includegraphics[width=\imagewidth]{\imagesource}};
            \makerectanglecorners{\imagename}{#2}{#3}{#4}{#5};
            \node[inner sep=0, below = 0.0cm of \imagename, scale=\imagewidth/((#4-#2)*\imagewidth)] (zoomcrop)  {
                \begin{tikzpicture}[inner sep=0]
                    \clip (firstcorner) rectangle (secondcorner);
                    \node[inner sep=0, anchor=south west] {\includegraphics[width=\imagewidth]{\imagesource}};
                \end{tikzpicture}
            };
        \end{tikzpicture}
    \end{subfigure}
}

%% file: figures/intro_example.tex
\begin{figure}[h]
  \center
  \begin{tikzpicture}[anchor=south west]
  \def\imagewidth{0.5\textwidth}
  \def\imagesource{images/introduction/father_children_on_shoulder.jpg}
  \def\imagename{image_intro2}

  \node[name=\imagename, inner sep=0]{\includegraphics[width=\imagewidth]{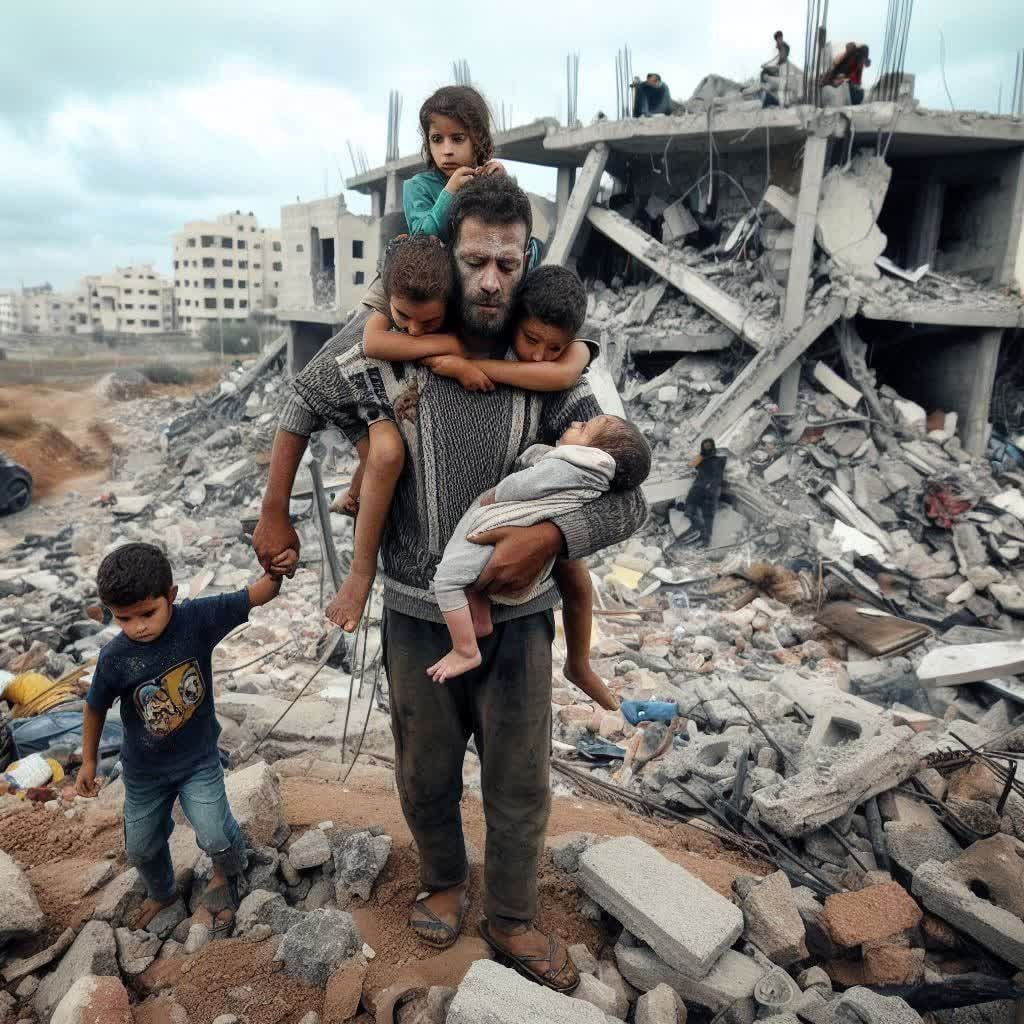}};

  \makerectanglecorners{\imagename}{0.39}{0.61}{0.5}{0.66};
  \drawbbox{firstcorner}{secondcorner}{\imagesource}{\imagewidth};

  \makeandplacecrop{\imagename}{1.1}{0.8}{\imagesource}{\imagewidth}{3};
  \node[below =0cm of cropbox, align=center] (caption) {Merged hands};

  \makebboxline{bbox.north}{cropbox.west}

  \makerectanglecorners{\imagename}{0.408}{0.327}{0.48}{0.37};
  \drawbbox{firstcorner}{secondcorner}{\imagesource}{\imagewidth};

  \makeandplacecrop{\imagename}{1.1}{0.1}{\imagesource}{\imagewidth}{3};
  \node[below =0cm of cropbox, align=center] (caption) {Too few toes};

  \makebboxline{bbox.east}{cropbox.west}

  \makerectanglecorners{\imagename}{0.3}{0.488}{0.4}{0.617};
  \drawbbox{firstcorner}{secondcorner}{\imagesource}{\imagewidth};

  \makeandplacecrop{\imagename}{-0.3}{0.7}{\imagesource}{\imagewidth}{2};
  \node[below =0cm of cropbox, align=center] (caption) {Extra limb};

  \makebboxline{bbox.west}{cropbox.east}

  \makerectanglecorners{\imagename}{0.097}{0.082}{0.26}{0.163};
  \drawbbox{firstcorner}{secondcorner}{\imagesource}{\imagewidth};

  \makeandplacecrop{\imagename}{-0.4}{0.1}{\imagesource}{\imagewidth}{2};
  \node[below =0cm of cropbox, align=center] (caption) {Indistinct feet};

  \makebboxline{bbox.west}{cropbox.east}

  \makerectanglecorners{\imagename}{0.313}{0.374}{0.3720}{0.423};
  \drawbbox{firstcorner}{secondcorner}{\imagesource}{\imagewidth};

  \makeandplacecrop{\imagename}{-0.35}{0.4}{\imagesource}{\imagewidth}{4};
  \node[below =0cm of cropbox, align=center] (caption) {Number of toes};

  \makebboxline{bbox.west}{cropbox.east}

  \makerectanglecorners{\imagename}{0.3369}{0.5849}{0.421}{0.645};
  \drawbbox{firstcorner}{secondcorner}{\imagesource}{\imagewidth};

  \makeandplacecrop{\imagename}{1.1}{0.4}{\imagesource}{\imagewidth}{3};

  \node[below =0cm of cropbox, align=center] (caption) {Clothing Patterns};

  \makebboxline{bbox.east}{cropbox.west}

  \end{tikzpicture}

  \caption{Synthetic photograph with annotated artifacts: The hands of the two children around the neck seem merged. The baby's foot is missing toes, the boy who walks has indistinct feet, and there is an extra leg / or leg growing from the right shoulder of the man. This example demonstrates a typical weakness of generative models: The accurate depiction of human anatomy and details. Image Source: \href{https://donya-e-eqtesad.com/\%D8\%A8\%D8\%AE\%D8\%B4-\%D8\%B3\%D8\%A7\%DB\%8C\%D8\%AA-\%D8\%AE\%D9\%88\%D8\%A7\%D9\%86-62/4014731-\%D8\%A7\%DB\%8C\%D9\%86-\%D8\%B9\%DA\%A9\%D8\%B3-\%D8\%AF\%D9\%86\%DB\%8C\%D8\%A7-\%D8\%B1\%D8\%A7-\%D8\%AA\%DA\%A9\%D8\%A7\%D9\%86-\%D8\%AF\%D8\%A7\%D8\%AF-\%D8\%AA\%D8\%B5\%D9\%88\%DB\%8C\%D8\%B1}{Link}}
  \label{fig:intro_example}
\end{figure}
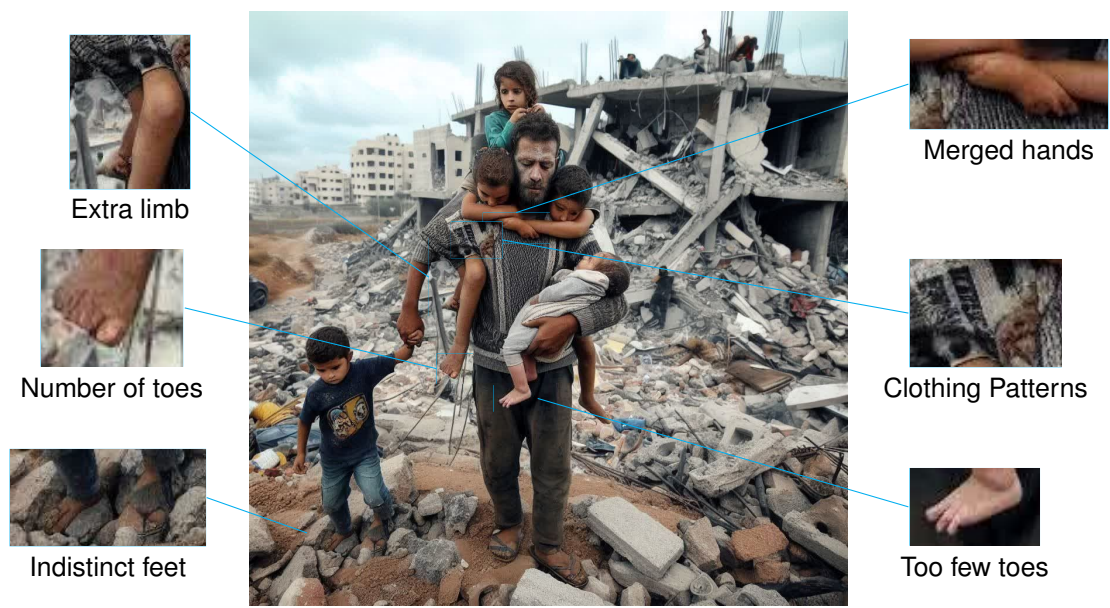

%% file: figures/artifacts_one_of_each.tex
\begin{figure}[ht]
    \centering
    \begin{subfigure}[b]{0.3\textwidth}
        \centering
        \includegraphics[width=\textwidth]{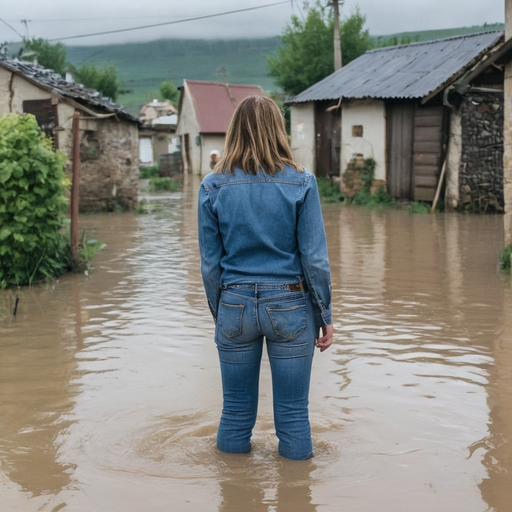}
        \caption{Physical}
    \end{subfigure}
    \quad
    \begin{subfigure}[b]{0.3\textwidth}
        \centering
        \includegraphics[width=\textwidth]{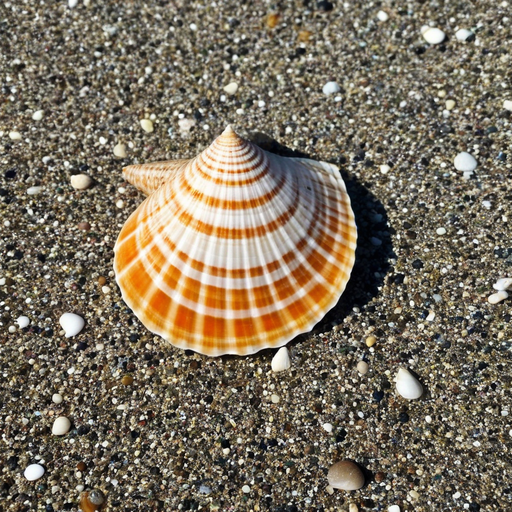}
        \caption{Geometrical}
    \end{subfigure}
    \quad
    \begin{subfigure}[b]{0.3\textwidth}
        \centering
        \includegraphics[width=\textwidth]{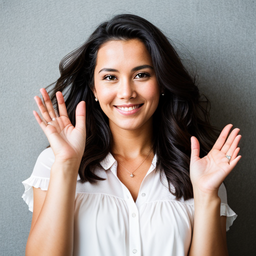}
        \caption{Human Anatomy}
    \end{subfigure}
    
    \vspace{\baselineskip}
    
    \begin{subfigure}[b]{0.3\textwidth}
        \centering
        \includegraphics[width=\textwidth]{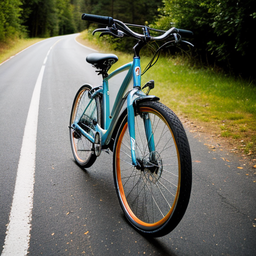}
        \caption{Semantic}
    \end{subfigure}
    \quad
    \begin{subfigure}[b]{0.3\textwidth}
        \centering
        \includegraphics[width=\textwidth]{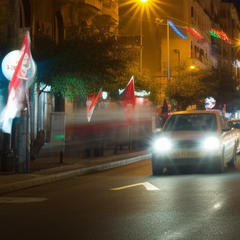}
        \caption{Distortion}
        \label{fig:artifact_overview:distortion}
    \end{subfigure}
    \quad
    \begin{subfigure}[b]{0.3\textwidth}
        \centering
        \includegraphics[width=\textwidth]{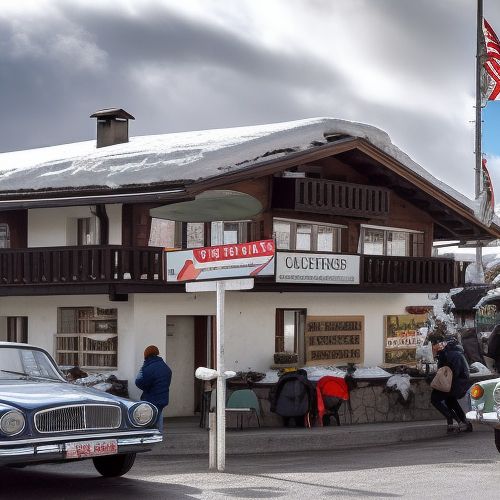}
        \caption{Text}
        \label{fig:artifact_overview:text}
    \end{subfigure}
    
    \caption{Synthetic image of each artifact category. \ref{fig:artifact_overview:distortion} and \ref{fig:artifact_overview:text} are from the Synthbuster dataset \cite{bammey_synthbuster_2023}, created with Midjourney-V5.}
    \label{fig:artifact_overview}
\end{figure}

%% file: figures/artifacts_overview.tex
\newcommand{\overviewimagewidth}{1.5cm}
\newcommand{\thickbboxbordersize}{very thick}
\newcommand{\formatcaptions}[1]{
    \footnotesize\textcolor{black}{#1}
}
\newcommand{\formatcaptionsubcategory}[1]{
    \footnotesize#1
}

\begin{figure}
	\centering
	\begin{tikzpicture}[
		align=center, inner sep=2.5pt, minimum height=20pt, minimum width=50pt, , text depth = 1.5,font = {\scriptsize}
		]
		\def\childdistance{3pt}
		\def\cminheight{30pt}
            \def\transparency{!5}
            \def\maincatsize{\normalsize}
            \def\subcatsize{\small}
		\node[] (link1) {};

		\node[fill=orange\transparency, rounded corners, minimum height= \cminheight, below=0cm of link1, font=\maincatsize] (human) {Human\\Anatomy};
		\node[fill=gray\transparency, rounded corners,  minimum height= \cminheight, right=\childdistance of human, font=\maincatsize] (geometrical) {Geometrical};
		\node[fill=blue\transparency, rounded corners, minimum height= \cminheight,right=\childdistance of geometrical, font=\maincatsize] (physical) {Physical};
		\node[fill=yellow\transparency, rounded corners, minimum height= \cminheight,right=\childdistance of physical, font=\maincatsize] (distortions) {
			Distortions
	};
		\node[fill=green\transparency, rounded corners,minimum height= \cminheight,right=\childdistance of distortions, font=\maincatsize] (semantic) {
			Semantic
		};
		\node[fill=red\transparency, rounded corners, minimum height= \cminheight,right=\childdistance of semantic, font=\maincatsize] (text) {
			Text
		};

		\node[fill=red\transparency, rounded corners, minimum height= \cminheight,below=\childdistance of text] (textcontent) {
			\includegraphics[width=\overviewimagewidth]{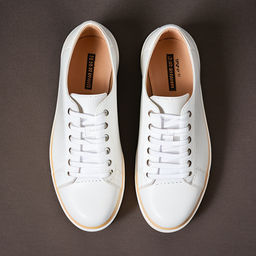}\\
                Unreadable\\
                labels\\
                inside shoes.
		};
		\draw (text.south) -| (textcontent.north);

		\node[fill=green\transparency, rounded corners, minimum height= \cminheight,below=\childdistance of semantic] (semanticcontent) {
			\includegraphics[width=\overviewimagewidth]{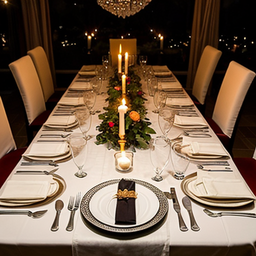}\\
                More table\\
                settings\\
                than chairs.
		};
		\draw (semantic.south) -| (semanticcontent.north);

		\node[fill=yellow\transparency, rounded corners, minimum height= \cminheight,below=\childdistance of distortions] (distortionscontent) {
			\includegraphics[width=\overviewimagewidth]{images/overview/one_of_each/distortion_3.png}\\
                Implausible\\
                motion\\
                blur.
		};
		\draw (distortions.south) -| (distortionscontent.north);

		\node[below=2.5cm of physical] (link2) {};
		\draw (physical.south) -| (link2.south);

		\node[fill=blue\transparency, rounded corners, minimum height= \cminheight,below left = 0.25cm and -1.1cm of link2, font=\subcatsize] (lighting) {Lighting};
		\draw (link2.south) -| (lighting.north);
  		\node[fill=blue\transparency, rounded corners, minimum height= \cminheight,below=\childdistance of lighting] (lightingcontent) {
			\includegraphics[width=\overviewimagewidth]{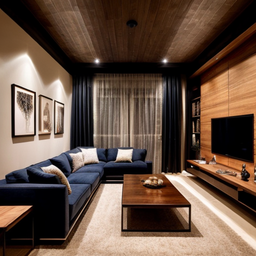} \\
            Light without\\
            source.
		};
		\draw (lighting.south) -| (lightingcontent.north);

		\node[fill=blue\transparency, rounded corners, minimum height= \cminheight,right=\childdistance of lighting, font=\subcatsize] (shadow) {Shadow};
		\draw (link2.south) -| (shadow.north);
    		\node[fill=blue\transparency, rounded corners, minimum height= \cminheight,below=\childdistance of shadow] (shadowcontent) {
			\includegraphics[width=\overviewimagewidth]{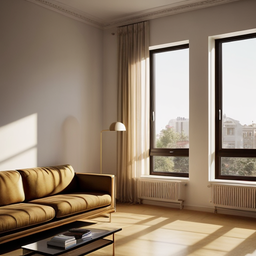}\\
            Window not\\
            matching its\\
            shadow.
		};
		\draw (shadow.south) -| (shadowcontent.north);

		\node[fill=blue\transparency, rounded corners, minimum height= \cminheight,right=\childdistance of shadow, font=\subcatsize] (reflection) {Reflection};
		\draw (link2.south) -| (reflection.north);
      		\node[fill=blue\transparency, rounded corners, minimum height= \cminheight,below=\childdistance of reflection] (reflectioncontent) {
			\includegraphics[width=\overviewimagewidth]{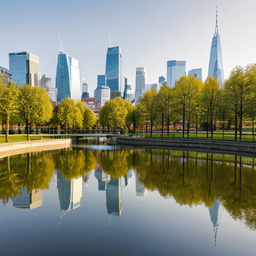}\\
            Buildings\\
            missing\\
            in reflection.
		};
		\draw (reflection.south) -| (reflectioncontent.north);

		\node[fill=blue\transparency, rounded corners, minimum height= \cminheight,right=\childdistance of reflection, font=\subcatsize] (gravity) {Gravity};
		\draw (link2.south) -| (gravity.north);
        	\node[fill=blue\transparency, rounded corners, minimum height= \cminheight,below=\childdistance of gravity] (gravitycontent) {
			\includegraphics[width=\overviewimagewidth]{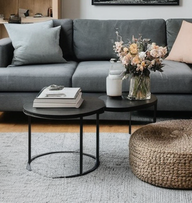}\\
            Table\\
            floating\\
            in the air.
		};
		\draw (gravity.south) -| (gravitycontent.north);

		\node[fill=blue\transparency, rounded corners, minimum height= \cminheight,right=\childdistance of gravity, font=\subcatsize] (other) {Other};
		\draw (link2.south) -| (other.north);
        		\node[fill=blue\transparency, rounded corners, minimum height= \cminheight,below=\childdistance of other] (othercontent) {
			\includegraphics[width=\overviewimagewidth]{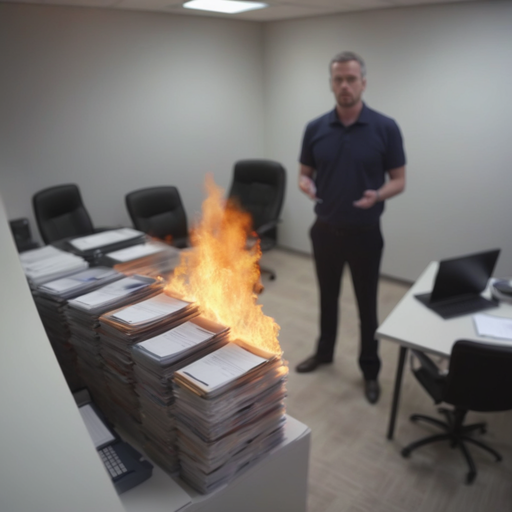}\\
            Fire has\\
            no effect.
		};
		\draw (other.south) -| (othercontent.north);

		\node[below=7cm of geometrical] (link3) {};
		\draw (geometrical.south) -| (link3.south);

		\node[fill=gray\transparency, rounded corners, minimum height= \cminheight,below left = 0.5cm and -1.1cm of link3, font=\subcatsize] (symmetry) {Symmetry};
		\draw (link3.south) -| (symmetry.north);
          		\node[fill=gray\transparency, rounded corners, minimum height= \cminheight,below=\childdistance of symmetry] (symmetrycontent) {
			\includegraphics[width=\overviewimagewidth]{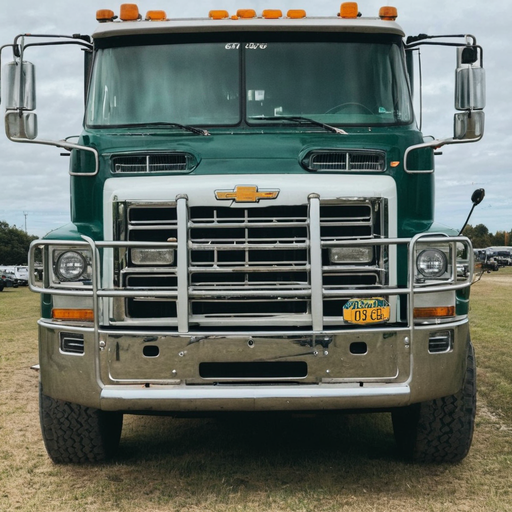}\\
            Asymmetries\\
            between left\\
            and right.
		};
		\draw (symmetry.south) -| (symmetrycontent.north);

		\node[fill=gray\transparency, rounded corners, minimum height= \cminheight,right=\childdistance of symmetry, font=\subcatsize] (shape) {Shape};
		\draw (link3.south) -| (shape.north);
            		\node[fill=gray\transparency, rounded corners, minimum height= \cminheight,below=\childdistance of shape] (shapecontent) {
			\includegraphics[width=\overviewimagewidth]{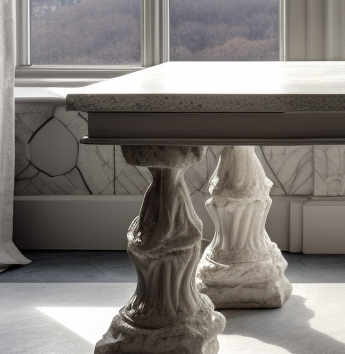}\\
            Strange,\\
            wobbly lines\\
            and shapes.
		};
		\draw (shape.south) -| (shapecontent.north);

		\node[fill=gray\transparency, rounded corners, minimum height= \cminheight,right=\childdistance of shape, font=\subcatsize] (scale) {Scale};
		\draw (link3.south) -| (scale.north);
  \node[fill=gray\transparency, rounded corners, minimum height= \cminheight,below=\childdistance of scale] (scalecontent) {
			\includegraphics[width=\overviewimagewidth]{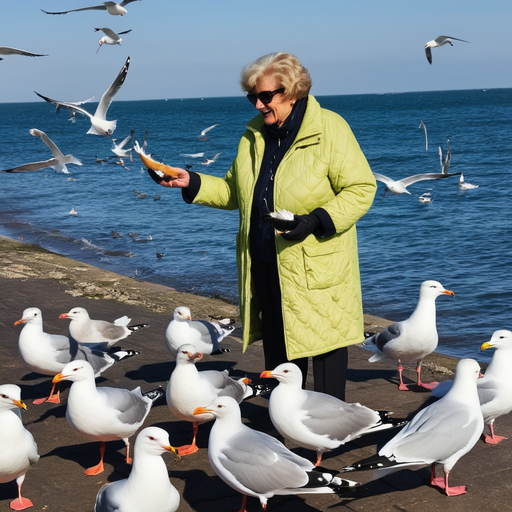}\\
               Birds too\\
               large, person\\
               too small.
		};
		\draw (scale.south) -| (scalecontent.north);

		\node[fill=gray\transparency, rounded corners, minimum height= \cminheight,right=\childdistance of scale, font=\subcatsize] (uniformity) {Uniformity};
		\draw (link3.south) -| (uniformity.north);
    \node[fill=gray\transparency, rounded corners, minimum height= \cminheight,below=\childdistance of uniformity] (uniformitycontent) {
			\includegraphics[width=\overviewimagewidth]{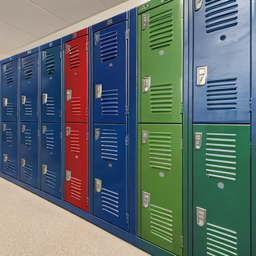}\\
   Inconsistent\\
   design and\\
   details.
		};
		\draw (uniformity.south) -| (uniformitycontent.north);

		\node[fill=gray\transparency, rounded corners, minimum height= \cminheight,right=\childdistance of uniformity, font=\subcatsize] (perspective) {Perspective};
		\draw (link3.south) -| (perspective.north);
    \node[fill=gray\transparency, rounded corners, minimum height= \cminheight,below=\childdistance of perspective] (perspectivecontent) {
			\includegraphics[width=\overviewimagewidth]{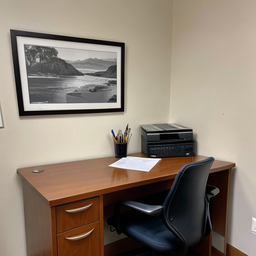}\\
            Lines not\\
            parallel or\\
            perpendicular.
		};
		\draw (perspective.south) -| (perspectivecontent.north);

		\node[fill=gray\transparency, rounded corners, minimum height= \cminheight,right=\childdistance of perspective, font=\subcatsize] (depth) {Depth};
		\draw (link3.south) -| (depth.north);
    \node[fill=gray\transparency, rounded corners, minimum height= \cminheight,below=\childdistance of depth] (depthcontent) {
			\includegraphics[width=\overviewimagewidth]{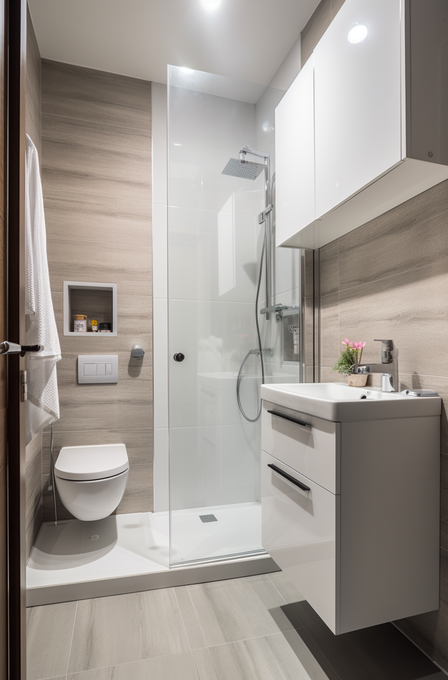}\\
            Inconsistent\\
            distance\\
            between\\
            glass and wall.
		};
		\draw (depth.south) -| (depthcontent.north);

  		\node[below=12.3cm of human] (link4) {};
		\draw (human.south) -| (link4.south);

		\node[fill=orange\transparency, rounded corners, minimum height= \cminheight,below left = 0.5cm and -1.5cm of link4, font=\subcatsize] (eyes) {Eyes};
		\draw (link4.south) -| (eyes.north);
    \node[fill=orange\transparency, rounded corners, minimum height= \cminheight,below=\childdistance of eyes] (eyescontent) {
			\includegraphics[width=\overviewimagewidth]{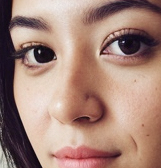}\\
            Unequal\\
            size and\\
            irregular\\
            shape.
		};
		\draw (eyes.south) -| (eyescontent.north);

		\node[fill=orange\transparency, rounded corners, minimum height= \cminheight,right=\childdistance of eyes, font=\subcatsize] (teeth) {Teeth};
		\draw (link4.south) -| (teeth.north);
    \node[fill=orange\transparency, rounded corners, minimum height= \cminheight,below=\childdistance of teeth] (teethcontent) {
			\includegraphics[width=\overviewimagewidth]{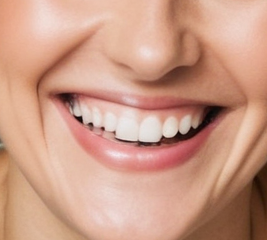}\\
            Asymmetrical\\
            shape and\\
            number.
		};
		\draw (teeth.south) -| (teethcontent.north);

		\node[fill=orange\transparency, rounded corners, minimum height= \cminheight,right=\childdistance of teeth, font=\subcatsize] (hair) {Hair};
		\draw (link4.south) -| (hair.north);
      \node[fill=orange\transparency, rounded corners, minimum height= \cminheight,below=\childdistance of hair] (haircontent) {
			\includegraphics[width=\overviewimagewidth]{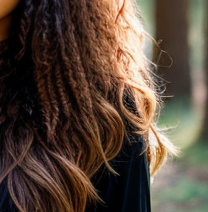}\\
            Inconsistent\\
            texture.
		};
		\draw (hair.south) -| (haircontent.north);

		\node[fill=orange\transparency, rounded corners, minimum height= \cminheight,right=\childdistance of hair, font=\subcatsize] (hands) {Hands};
		\draw (link4.south) -| (hands.north);
        \node[fill=orange\transparency, rounded corners, minimum height= \cminheight,below=\childdistance of hands] (handscontent) {
			\includegraphics[width=\overviewimagewidth]{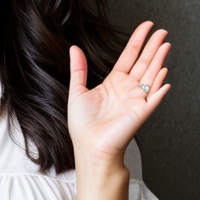}\\
            Inconsistent\\
            finger lengths,\\
            too many\\
            fingers.
		};
		\draw (hands.south) -| (handscontent.north);

		\node[fill=orange\transparency, rounded corners, minimum height= \cminheight,right=\childdistance of hands, font=\subcatsize] (body) {Body\\Shape};
		\draw (link4.south) -| (body.north);
          \node[fill=orange\transparency, rounded corners, minimum height= \cminheight,below=\childdistance of body] (bodycontent) {
			\includegraphics[width=\overviewimagewidth]{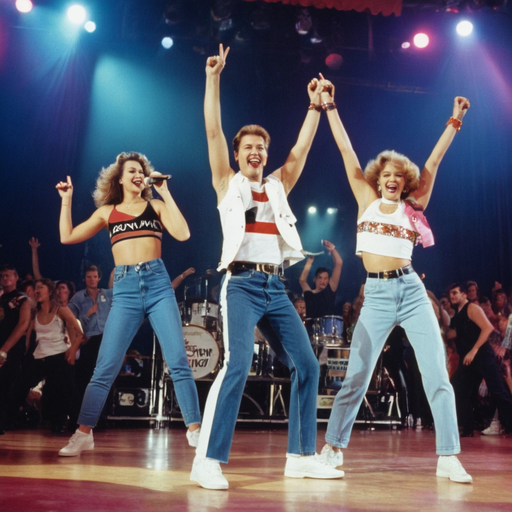}\\
            Unequal\\
            arm lengths,\\
            deformed\\
            people in\\
            background.
		};
		\draw (body.south) -| (bodycontent.north);

		\node[fill=orange\transparency, rounded corners, minimum height= \cminheight,right=\childdistance of body, font=\subcatsize] (wearables) {Wearables};
		\draw (link4.south) -| (wearables.north);
            \node[fill=orange\transparency, rounded corners, minimum height= \cminheight,below=\childdistance of wearables] (wearablescontent) {
			\includegraphics[width=\overviewimagewidth]{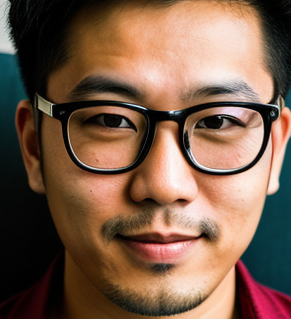}\\
            Asymmetrical\\
            glasses.
		};
		\draw (wearables.south) -| (wearablescontent.north);

	\end{tikzpicture}
\caption{Artifact taxonomy. Distortion, shape, and depth examples from Synthbuster \cite{bammey_synthbuster_2023}, generated with Midjourney v5.}
\label{fig:artifact_taxonomy}
\end{figure}

%% file: figures/people_intro.tex
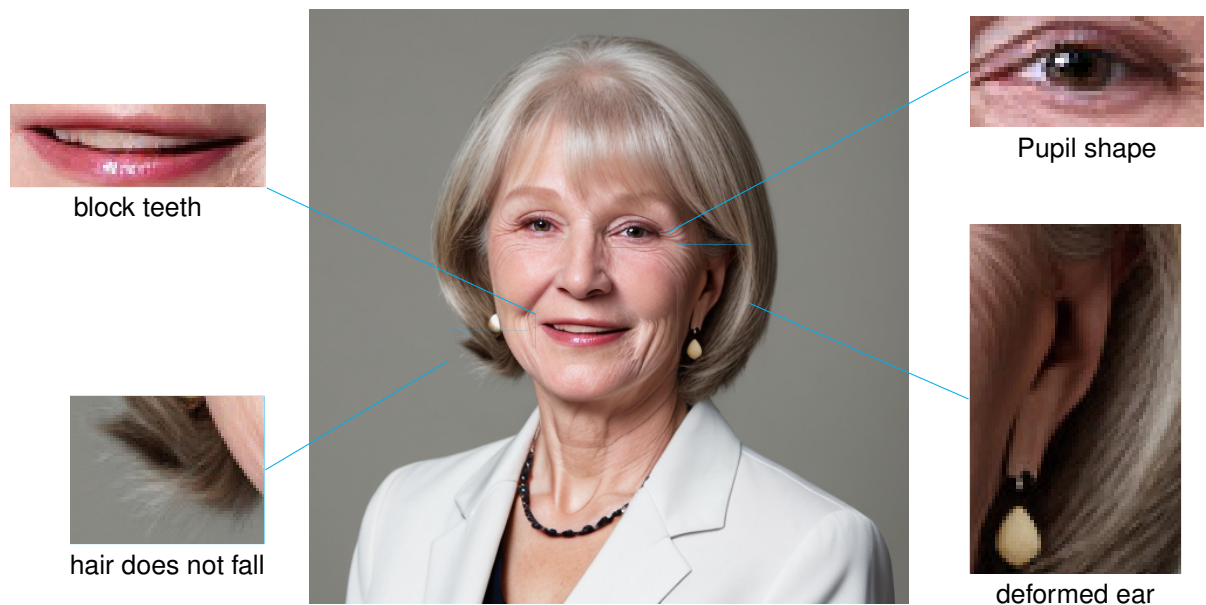
\begin{figure}[h]
\centering
\begin{tikzpicture}[anchor=south west]
\def\imagename{img}
\def\imagewidth{0.5\textwidth}
\def\imagesource{images/humans/linkedin/SD15_realisticVisionV60B1_v51VAE_13.png}
\node[name=\imagename, inner sep=0]{\includegraphics[width=\imagewidth]{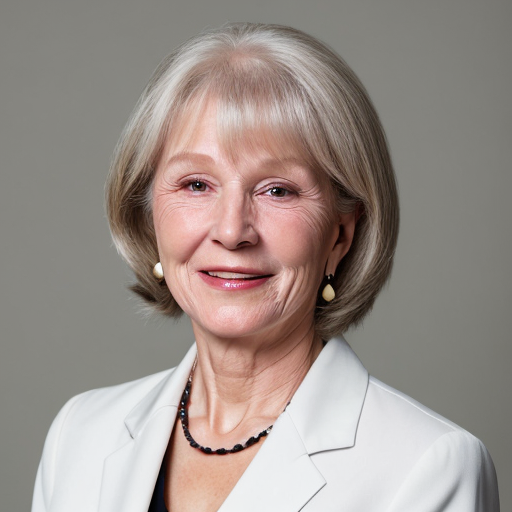}};

  \makerectanglecorners{\imagename}{0.232}{0.359}{0.361}{0.458};
  \drawbbox{firstcorner}{secondcorner}{\imagesource}{\imagewidth};

  \makeandplacecrop{\imagename}{-0.4}{0.1}{\imagesource}{\imagewidth}{2.5};

  \node[below =0cm of cropbox, align=center] (caption) {hair does not fall};

  \makebboxline{bbox.west}{cropbox.east}

  \makerectanglecorners{\imagename}{0.498}{0.60}{0.595}{0.646};
  \drawbbox{firstcorner}{secondcorner}{\imagesource}{\imagewidth};

  \makeandplacecrop{\imagename}{1.1}{0.8}{\imagesource}{\imagewidth}{4};

  \node[below =0cm of cropbox, align=center] (caption) {Pupil shape};

  \makebboxline{bbox.east}{cropbox.west}

  \makerectanglecorners{\imagename}{0.615}{0.408}{0.732}{0.603};
  \drawbbox{firstcorner}{secondcorner}{\imagesource}{\imagewidth};

  \makeandplacecrop{\imagename}{1.1}{0.05}{\imagesource}{\imagewidth}{3};

  \node[below =0cm of cropbox, align=center] (caption) {deformed ear};

  \makebboxline{bbox.east}{cropbox.west}

  \makerectanglecorners{\imagename}{0.378}{0.431}{0.548}{0.486};
  \drawbbox{firstcorner}{secondcorner}{\imagesource}{\imagewidth};

  \makeandplacecrop{\imagename}{-0.5}{0.7}{\imagesource}{\imagewidth}{2.5};

  \node[below =0cm of cropbox, align=center] (caption) {block teeth};

  \makebboxline{bbox.north west}{cropbox.south east}

\end{tikzpicture}
\caption{Synthetic portrait photography of a woman with small irregularities in the left eye, hair, teeth, and ear form. }
\label{fig:people_intro}
\end{figure}

%% file: figures/people_eyes.tex
\begin{figure}[ht]
\centering
  \begin{subfigure}[b]{0.2\textwidth}
      \begin{tikzpicture}[anchor=south west]
          \def\imagewidth{\textwidth}
          \def\imagesource{images/humans/linkedin/SD15_realisticVision_linkedin_1.png}
          \def\imagename{imagelinkedin1}

          \node[name=\imagename, inner sep=0]{\includegraphics[width=\imagewidth]{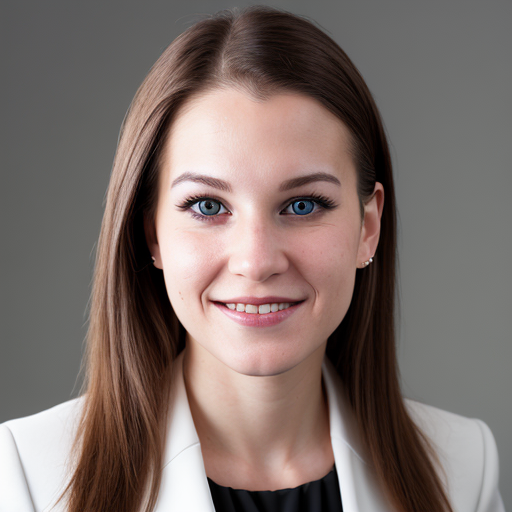}};
          \makerectanglecorners{\imagename}{0.3}{0.5}{0.7}{0.7};

           \node[inner sep=0, below = 0.0cm of \imagename, scale=\imagewidth/((0.7-0.3)*\imagewidth)] (zoomcrop)  {
                \begin{tikzpicture}[inner sep=0]
                    \clip(firstcorner) rectangle (secondcorner);
                  \node[inner sep=0, anchor=south west] {\includegraphics[width=\imagewidth]{\imagesource}};
                \end{tikzpicture}
                };
      \end{tikzpicture}
      \end{subfigure}
  \quad
    \begin{subfigure}[b]{0.2\textwidth}
        \begin{tikzpicture}[anchor=south west]
          \def\imagewidth{\textwidth}
          \def\imagesource{images/humans/linkedin/SD15_realisticVision_linkedin_2.png}
          \def\imagename{imagelinkedin2}

          \node[name=\imagename, inner sep=0]{\includegraphics[width=\imagewidth]{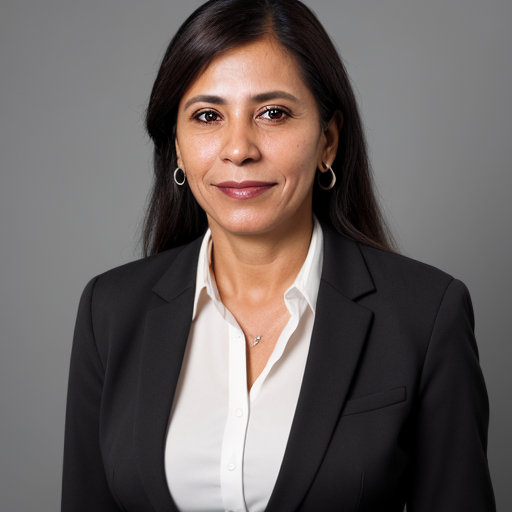}};
          \makerectanglecorners{\imagename}{0.35}{0.725}{0.6}{0.85};

           \node[inner sep=0, below = 0.0cm of \imagename, scale=\imagewidth/((0.6-0.35)*\imagewidth)] (zoomcrop)  {
                \begin{tikzpicture}[inner sep=0]
                    \clip(firstcorner) rectangle (secondcorner);
                  \node[inner sep=0, anchor=south west] {\includegraphics[width=\imagewidth]{\imagesource}};
                \end{tikzpicture}
                };
      \end{tikzpicture}
     \end{subfigure}
     \quad
    \begin{subfigure}[b]{0.2\textwidth}

        \begin{tikzpicture}[anchor=south west]
          \def\imagewidth{\textwidth}
          \def\imagesource{images/humans/linkedin/sdxl_juggernaut_xl_v8_linkedin_4.png}
          \def\imagename{imagelinkedin3}

          \node[name=\imagename, inner sep=0]{\includegraphics[width=\imagewidth]{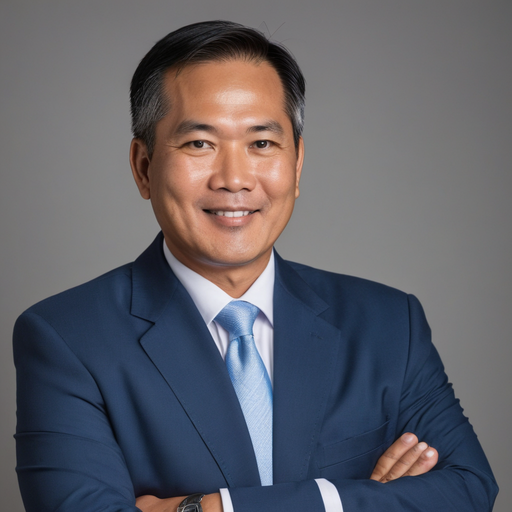}};
          \makerectanglecorners{\imagename}{0.325}{0.675}{0.575}{0.8};

           \node[inner sep=0, below = 0.0cm of \imagename, scale=\imagewidth/((0.575-0.325)*\imagewidth)] (zoomcrop)  {
                \begin{tikzpicture}[inner sep=0]
                    \clip(firstcorner) rectangle (secondcorner);
                  \node[inner sep=0, anchor=south west] {\includegraphics[width=\imagewidth]{\imagesource}};
                \end{tikzpicture}
                };
      \end{tikzpicture}
    \end{subfigure}
     \quad
    \begin{subfigure}[b]{0.2\textwidth}
        \centering
        Real
        \begin{tikzpicture}[anchor=south west]
          \def\imagewidth{\textwidth}
          \def\imagesource{images/humans/linkedin/real_portrait_cropped_1.jpg}
          \def\imagename{imagelinkedin4}

          \node[name=\imagename, inner sep=0]{\includegraphics[width=\imagewidth]{\imagesource}};
          \makerectanglecorners{\imagename}{0.3}{0.48}{0.68}{0.67};

           \node[inner sep=0, below = 0.0cm of \imagename, scale=\imagewidth/((0.68-0.3)*\imagewidth)] (zoomcrop)  {
                \begin{tikzpicture}[inner sep=0]
                    \clip(firstcorner) rectangle (secondcorner);
                  \node[inner sep=0, anchor=south west] {\includegraphics[width=\imagewidth]{\imagesource}};
                \end{tikzpicture}
                };
      \end{tikzpicture}
    \end{subfigure}

  \caption{Eye artifacts can best revealed by zooming in. The picture at the right shows a real photograph as a comparison (\href{https://unsplash.com/photos/a-woman-standing-in-front-of-a-sculpture-O3D-teBz0Bg}{Source: Philip White/Unsplash}).}

  \label{fig:people:eyes}
\end{figure}

%% file: figures/people_teeth.tex
\begin{figure}[h]
\centering

  \imagewithcropbelow{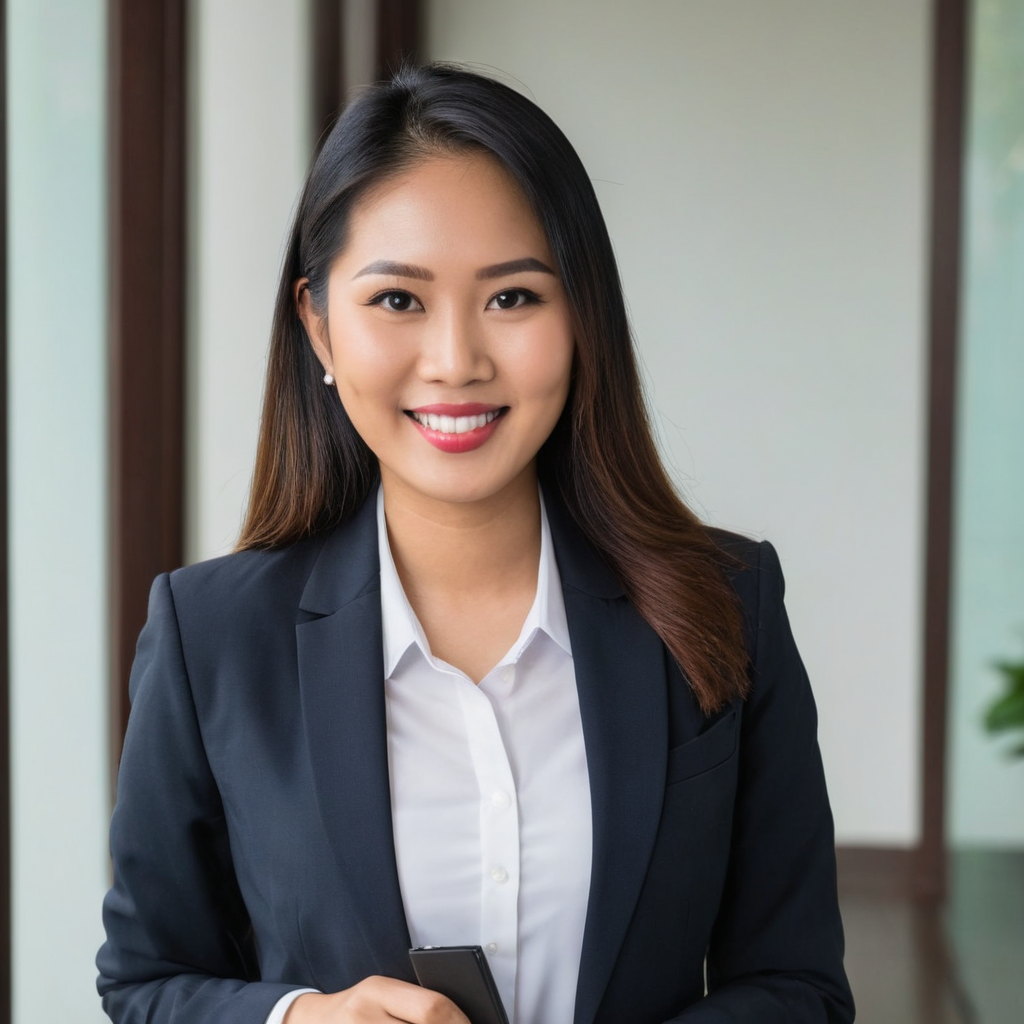}{0.4}{0.55}{0.5}{0.617}{0.2\textwidth}{}
  \quad
    \imagewithcropbelow{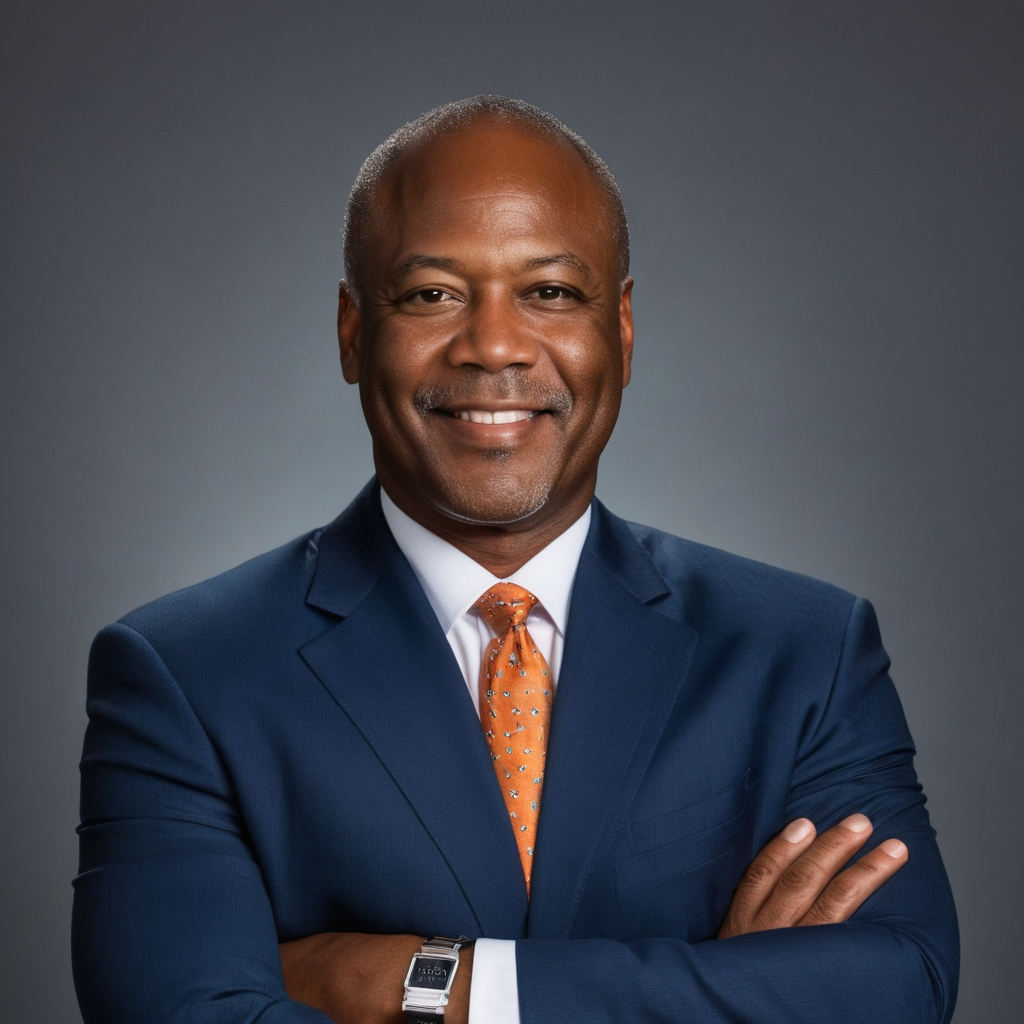}{0.4}{0.55}{0.55}{0.651}{0.2\textwidth}{}
  \quad
    \imagewithcropbelow{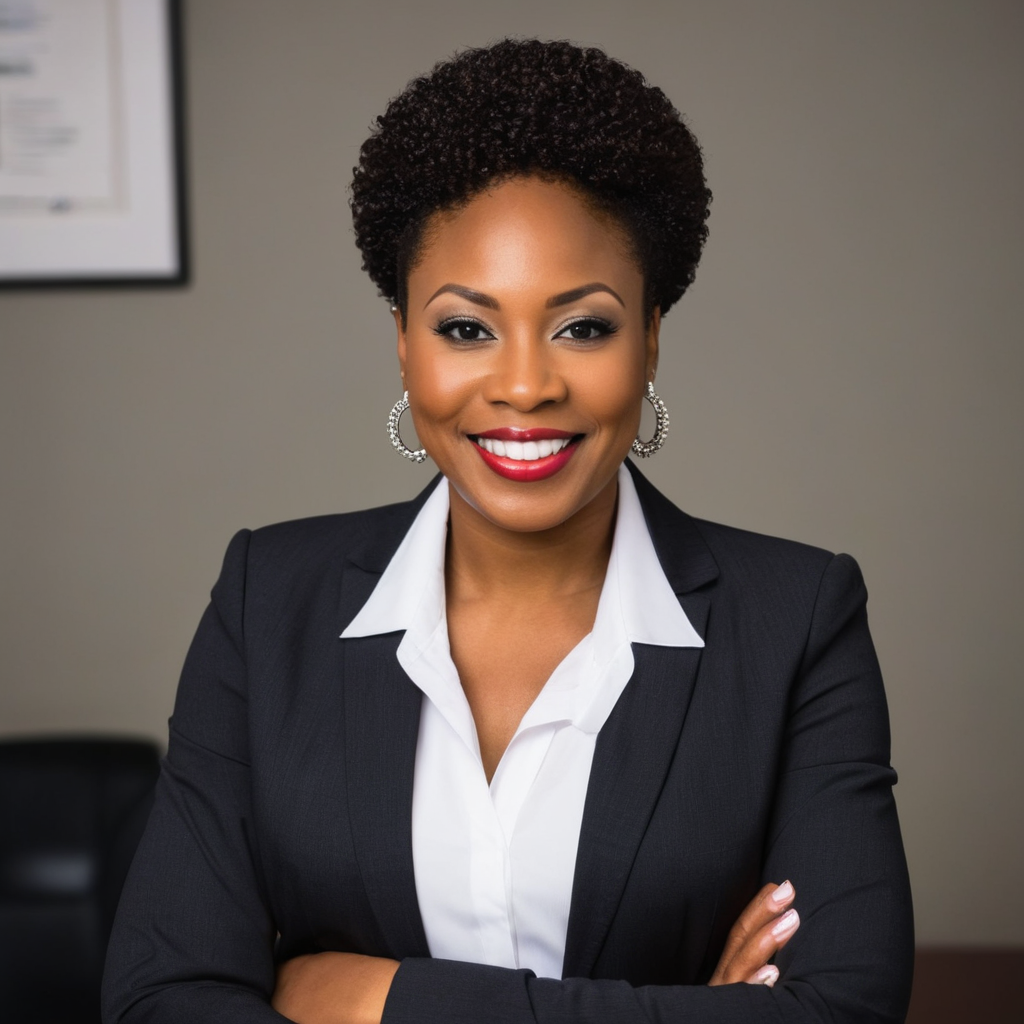}{0.45}{0.525}{0.575}{0.61}{0.2\textwidth}{}
  \quad
    \imagewithcropbelow{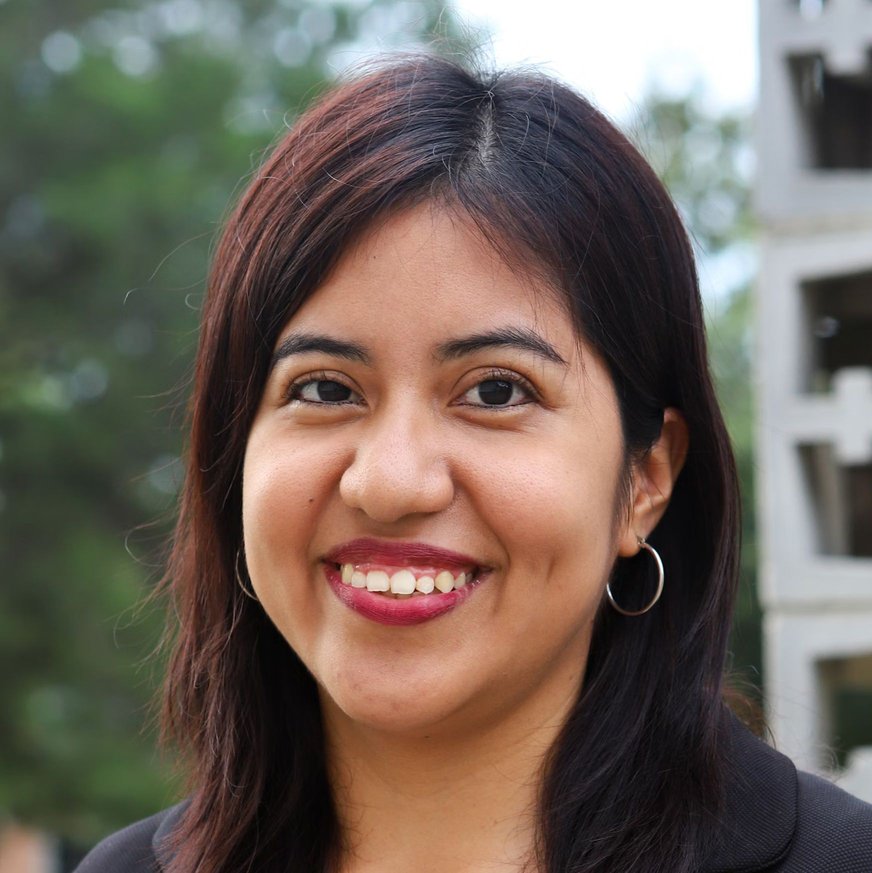}{0.33}{0.25}{0.59}{0.43}{0.2\textwidth}{Real}

  \caption{Synthetic photographs (first three from the left) with highlighted teeth. Teeth boundaries may not always be visible or may be omitted. As a comparison a real photograph (right), (\href{https://unsplash.com/photos/a-woman-standing-in-front-of-a-sculpture-O3D-teBz0Bg}{Source Philip White/Unsplash}).}
  \label{fig:people_teeth}
\end{figure}

%% file: figures/people_ears.tex
\begin{figure}[h]
\centering

  \imagewithcropbelow{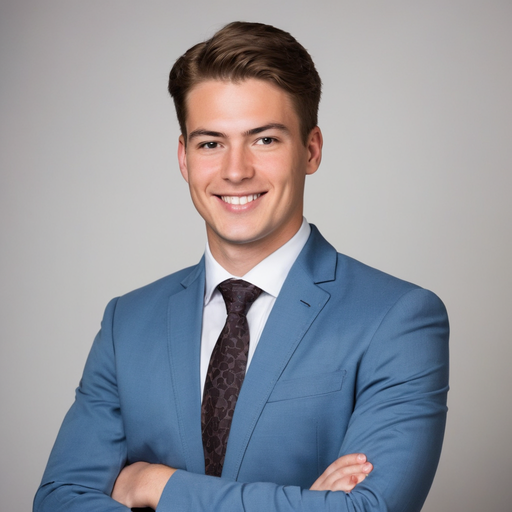}{0.34}{0.63}{0.65}{0.76}{0.2\textwidth}{}
  \quad
    \imagewithcropbelow{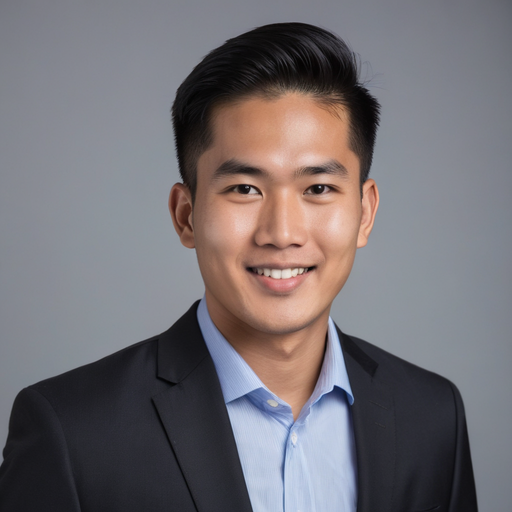}{0.31}{0.48}{0.77}{0.68}{0.2\textwidth}{}
  \quad
    \imagewithcropbelow{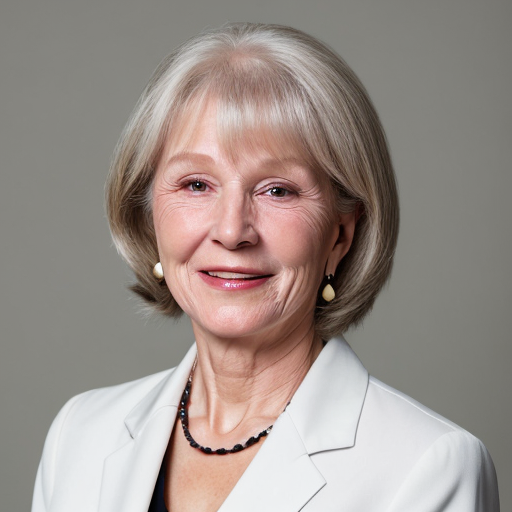}{0.265}{0.35}{0.69}{0.53}{0.2\textwidth}{}
  \quad
    \imagewithcropbelow{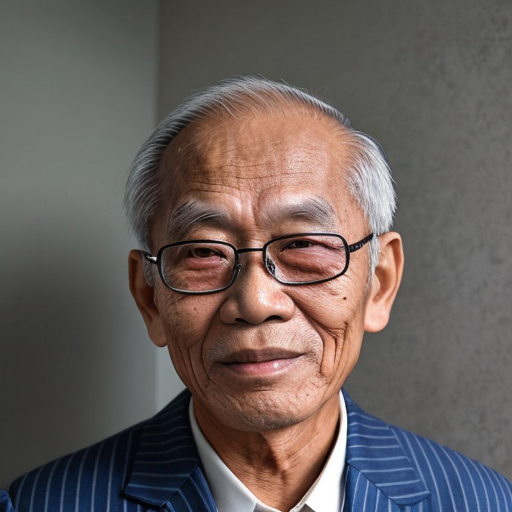}{0.246}{0.40}{0.73}{0.61}{0.2\textwidth}{}
  \caption{Synthetic photographs with asymmetric / malformed ears (first and second from left), and asymmetric earrings (third) and glasses (fourth).}
  \label{fig:people_ears_wearables}
\end{figure}

%% file: figures/people_hair.tex
\begin{figure}[!htb]
\centering

  \imagewithcropbelow{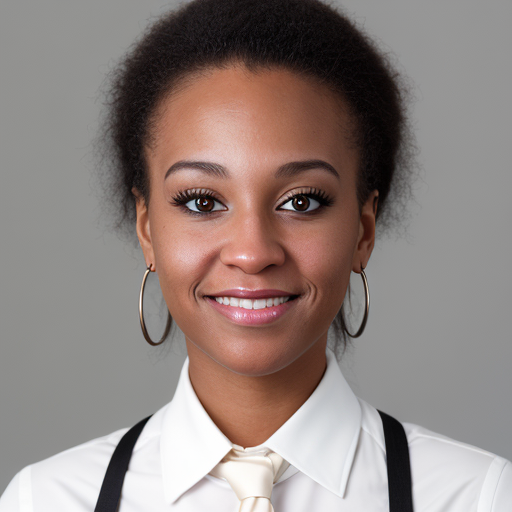}{0.16}{0.62}{0.79}{1}{0.2\textwidth}{}
  \quad
    \imagewithcropbelow{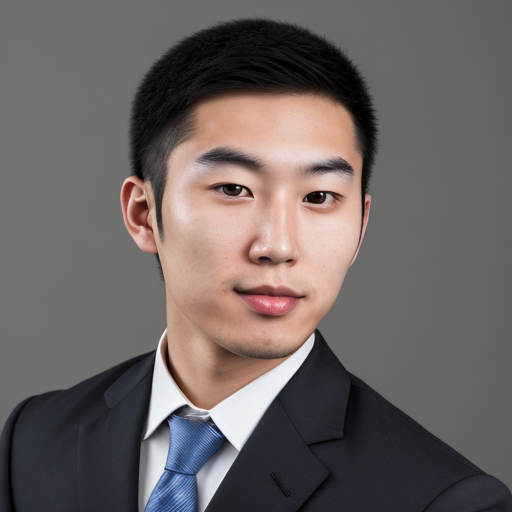}{0.246}{0.64}{0.69}{0.90}{0.2\textwidth}{}
  \quad
    \imagewithcropbelow{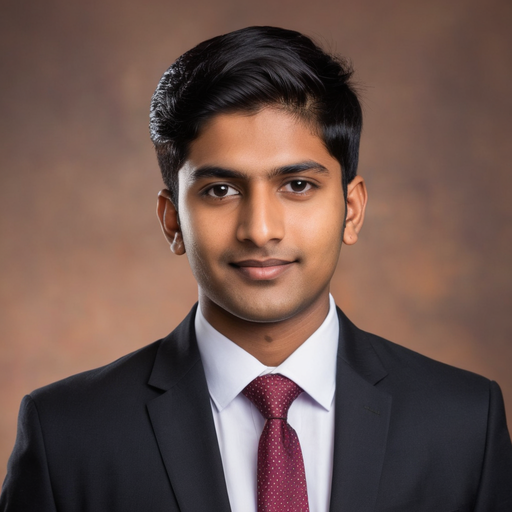}{0.303}{0.59}{0.74}{0.85}{0.2\textwidth}{}
  \quad
    \imagewithcropbelow{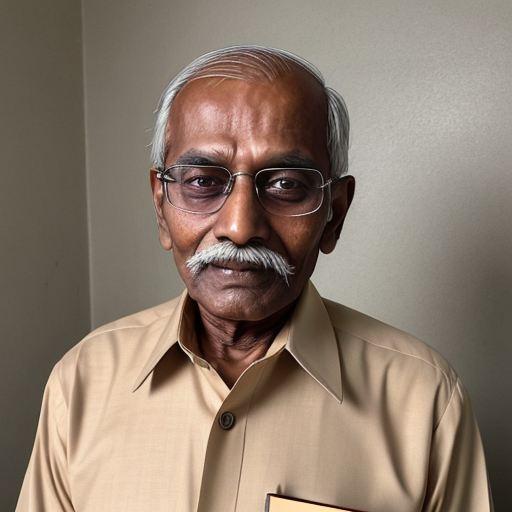}{0.359}{0.402}{0.607}{0.55}{0.2\textwidth}{}

  \caption{Synthetic photographs with hair artifacts. From left to right: blur and lack of detail, unnatural hair texture, asymmetric hair cut, and unnatural / paint-like texture.}
  \label{fig:people_hair}
\end{figure}

%% file: figures/people_groupe_single.tex
\begin{figure}[!htb]
\centering
\begin{tikzpicture}[anchor=south west]
\def\imagename{img}
\def\imagewidth{0.4\textwidth}
\def\imagesource{images/humans/refugees/sdxl_juggernaut_xl_v8_7.png}
\node[name=\imagename, inner sep=0]{\includegraphics[width=\imagewidth]{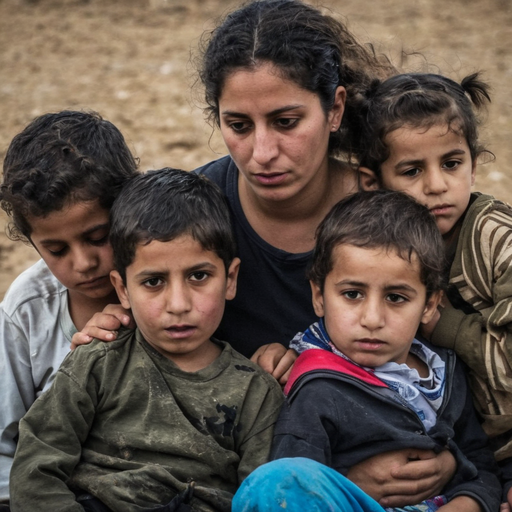}};

\makerectanglecorners{\imagename}{0.125}{0.325}{0.275}{0.425}
\drawbbox{firstcorner}{secondcorner}{\imagesource}{\imagewidth}
\makeandplacecrop{\imagename}{-0.4}{0.1}{\imagesource}{\imagewidth}{2}
\node[below =0cm of cropbox, align=center] (caption2) {deformed hands};
\makebboxline{bbox.west}{cropbox.north east}

\makerectanglecorners{\imagename}{0.692}{0.61}{0.749}{0.68}
\drawbbox{firstcorner}{secondcorner}{\imagesource}{\imagewidth}
\makeandplacecrop{\imagename}{-0.4}{0.7}{\imagesource}{\imagewidth}{4}
\node[below =0cm of cropbox, align=center] (caption2) {ear malformed};
\makebboxline{bbox.west}{cropbox.east}

\makerectanglecorners{\imagename}{0.65}{0.4}{0.825}{0.45}
\drawbbox{firstcorner}{secondcorner}{\imagesource}{\imagewidth}
\makeandplacecrop{\imagename}{-0.6}{0.4}{\imagesource}{\imagewidth}{3}
\node[below =0cm of cropbox, align=center] (caption5) {Eye asymmetry};
\makebboxline{bbox.west}{cropbox.north east}

\makerectanglecorners{\imagename}{0.6}{0.75}{0.8}{0.95}
\drawbbox{firstcorner}{secondcorner}{\imagesource}{\imagewidth}
\makeandplacecrop{\imagename}{1.1}{0.5}{\imagesource}{\imagewidth}{2.5}
\node[below =0cm of cropbox, align=center] (caption3) {Hair merged};
\makebboxline{bbox.east}{cropbox.west}

\makerectanglecorners{\imagename}{0.67}{0.009}{0.918}{0.14}
\drawbbox{firstcorner}{secondcorner}{\imagesource}{\imagewidth}
\makeandplacecrop{\imagename}{1.1}{0.1}{\imagesource}{\imagewidth}{2}
\node[below =0cm of cropbox, align=center] (caption5) {Fingers merged};
\makebboxline{bbox.east}{cropbox.west}

\end{tikzpicture}
\caption{Group of people with distinct artifacts related to hair, hands, and eyes.}
\label{fig:people_group_individual}
\end{figure}
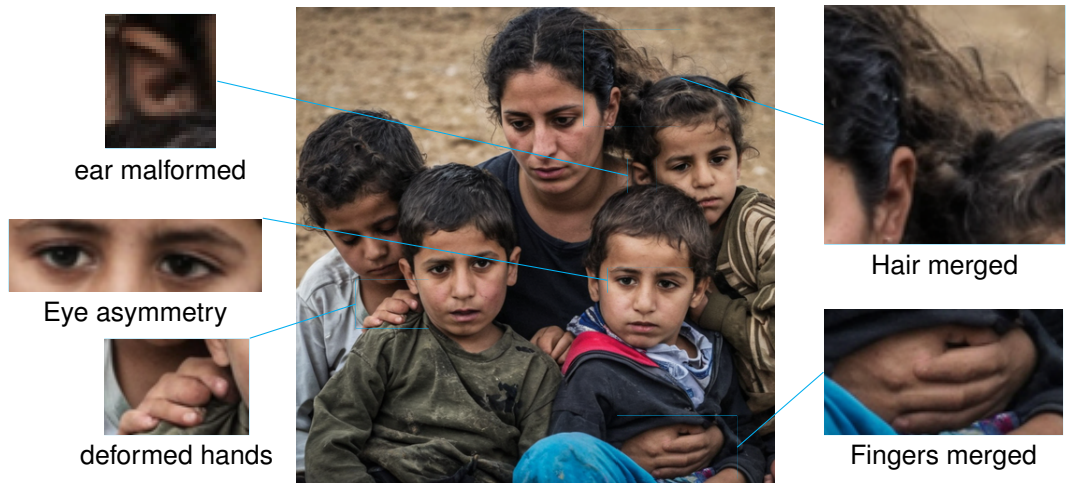

%% file: figures/people_group_war.tex
\begin{figure}[!htb]
\centering
    \begin{subfigure}[b]{0.3\textwidth}
        \begin{tikzpicture}[anchor=south west]
            \def\imagewidth{\textwidth}
            \def\imagesource{images/humans/refugees/sdxl_juggernaut_xl_v8_1.png}
            \node[name=img, inner sep=0]{\includegraphics[width=\imagewidth]{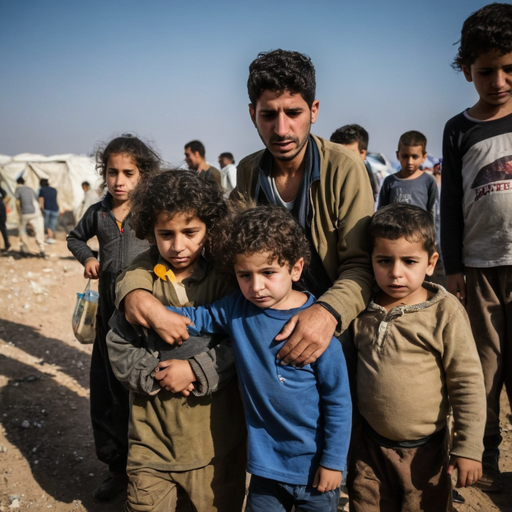}};
        \end{tikzpicture}
    \end{subfigure}
  \quad
    \begin{subfigure}[b]{0.3\textwidth}
        \begin{tikzpicture}[anchor=south west]
            \def\imagewidth{\textwidth}
            \def\imagesource{images/humans/refugees/sdxl_juggernaut_xl_v8_3.png}
            \node[name=img, inner sep=0]{\includegraphics[width=\imagewidth]{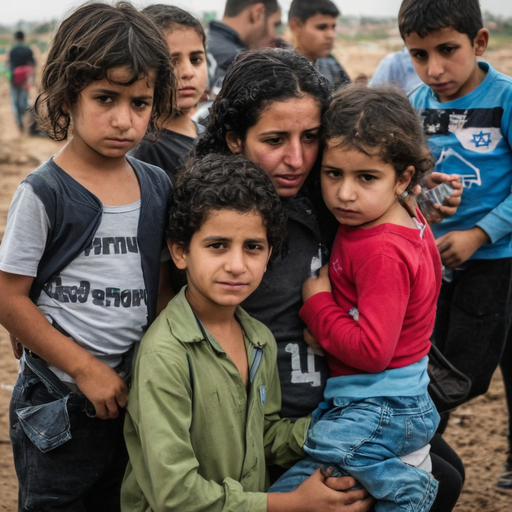}};
        \end{tikzpicture}
    \end{subfigure}
    \quad
    \begin{subfigure}[b]{0.3\textwidth}
        \begin{tikzpicture}[anchor=south west]
            \def\imagewidth{\textwidth}
            \def\imagesource{images/humans/refugees/sdxl_juggernaut_xl_v8_4.png}
            \node[name=img, inner sep=0]{\includegraphics[width=\imagewidth]{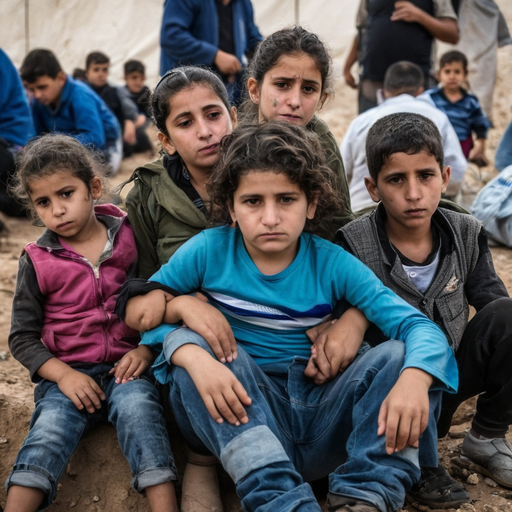}};
        \end{tikzpicture}
    \end{subfigure}

    \caption{Synthetic images of small groups of people in a refugee camp. Many artefacts related to incorrect human anatomy can be observed, particularly malformed hands.}
    \label{fig:people_refugees_group}
\end{figure}

%% file: figures/people_met_gala_gomez.tex
\begin{figure}[!htb]
\centering
\begin{tikzpicture}[anchor=south west]
\def\imagename{img}
\def\imagewidth{0.5\textwidth}
\def\imagesource{images/humans/met_gala_2024/selena_gomez_3.jpg}
\node[name=\imagename, inner sep=0]{\includegraphics[width=\imagewidth]{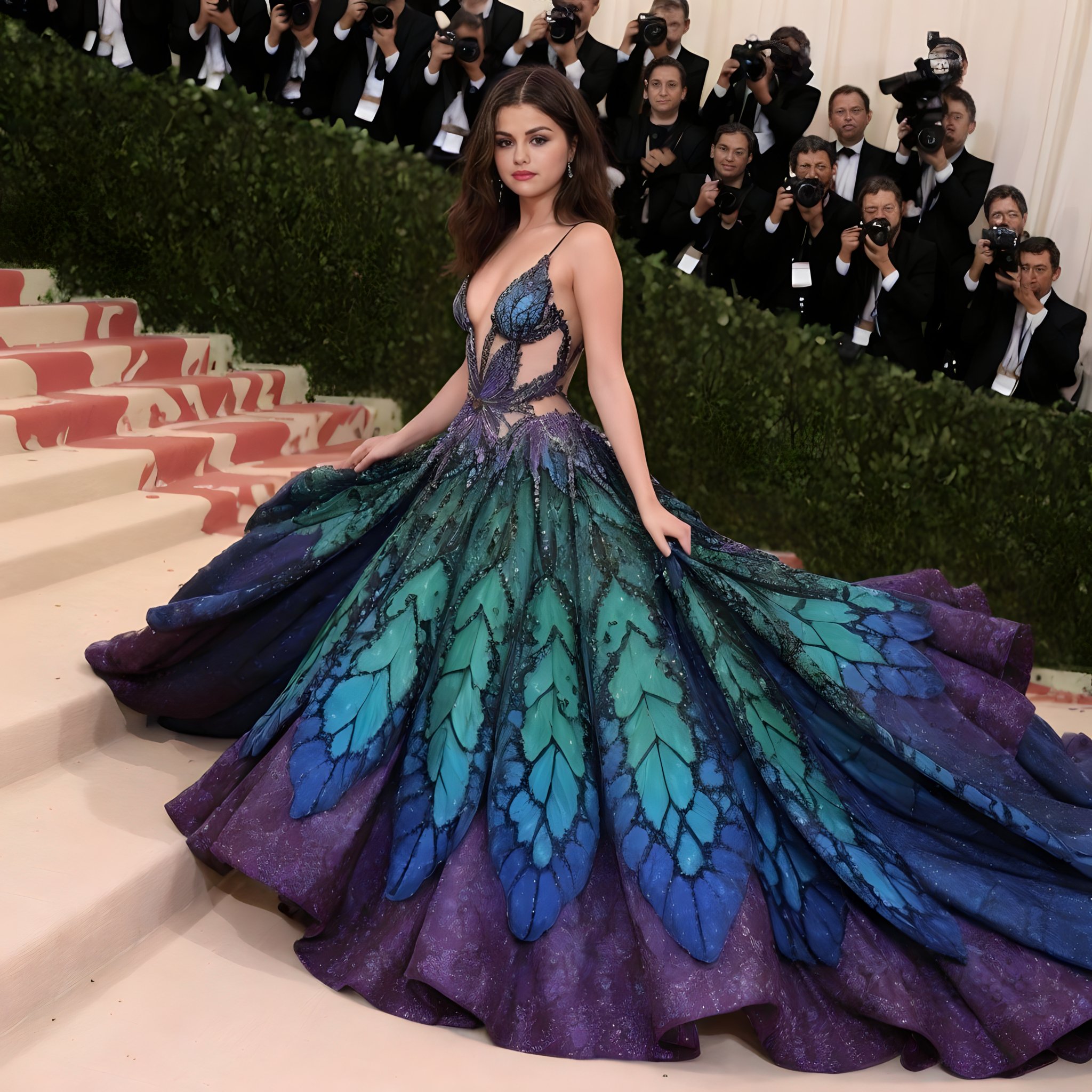}};

  \makerectanglecorners{\imagename}{0.89}{0.73}{0.99}{0.84};
  \drawbbox{firstcorner}{secondcorner}{\imagesource}{\imagewidth};

  \makeandplacecrop{\imagename}{1.1}{0.1}{\imagesource}{\imagewidth}{4};

  \node[below =0cm of cropbox, align=center] (caption) {Disfigurement \& Distortions};

  \makebboxline{bbox.south}{cropbox.west}

  \makerectanglecorners{\imagename}{0.70}{0.80}{0.78}{0.88};
  \drawbbox{firstcorner}{secondcorner}{\imagesource}{\imagewidth};

  \makeandplacecrop{\imagename}{1.1}{0.7}{\imagesource}{\imagewidth}{4};

  \node[below =0cm of cropbox, align=center] (caption) {Hands \& Camera Merge};

  \makebboxline{bbox.east}{cropbox.west}

  \makerectanglecorners{\imagename}{0.48}{0.95}{0.625}{1};
  \drawbbox{firstcorner}{secondcorner}{\imagesource}{\imagewidth};

  \makeandplacecrop{\imagename}{-0.5}{0.7}{\imagesource}{\imagewidth}{3};

  \node[below =0cm of cropbox, align=center] (caption) {Disfigured Faces};

  \makebboxline{bbox.west}{cropbox.east}

\end{tikzpicture}

\caption{Synthetic photography of Selena Gomez apparently taking part at the Met Gala. While she is depicted rather convincingly the photographers in the back lack detail or have malformed cameras ( \href{https://x.com/Leonaderune/status/1787708297627533399/photo/3}{Source: Twitter/X}.)}
\label{fig:people_meta_gala_gomez}
\end{figure}
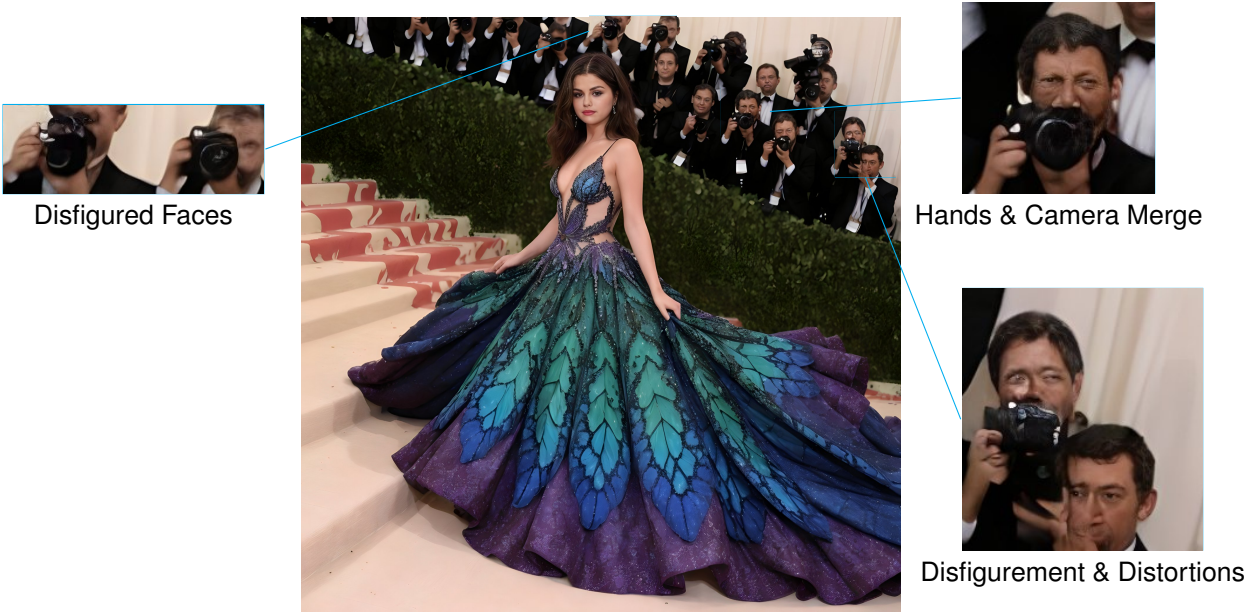

%% file: figures/indoors_example.tex
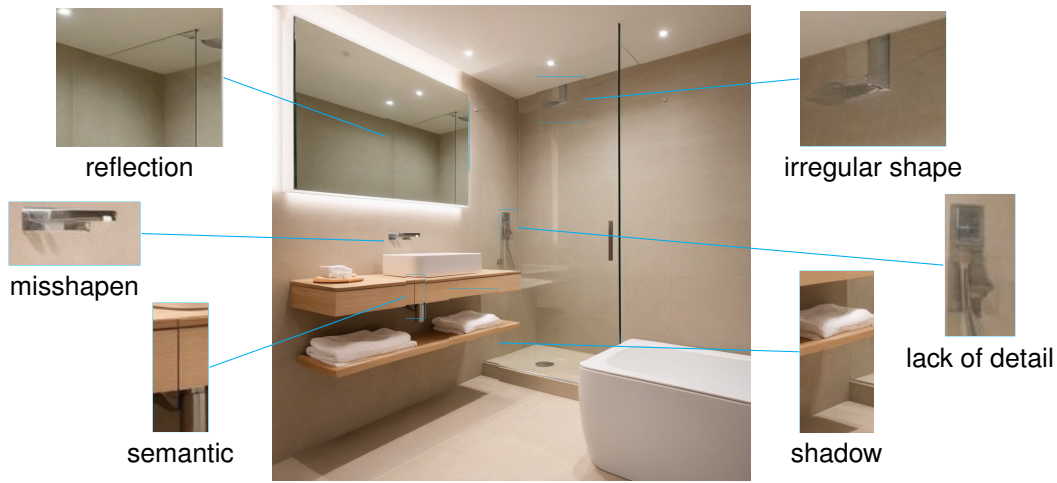
\begin{figure}[h]
  \center
  \begin{tikzpicture}[anchor=south west]
  \def\imagewidth{0.4\textwidth}
  \def\imagesource{images/indoors/apartments/detailed_example_2.png}
  \def\imagename{img}

  \node[name=\imagename, inner sep=0]{\includegraphics[width=\imagewidth]{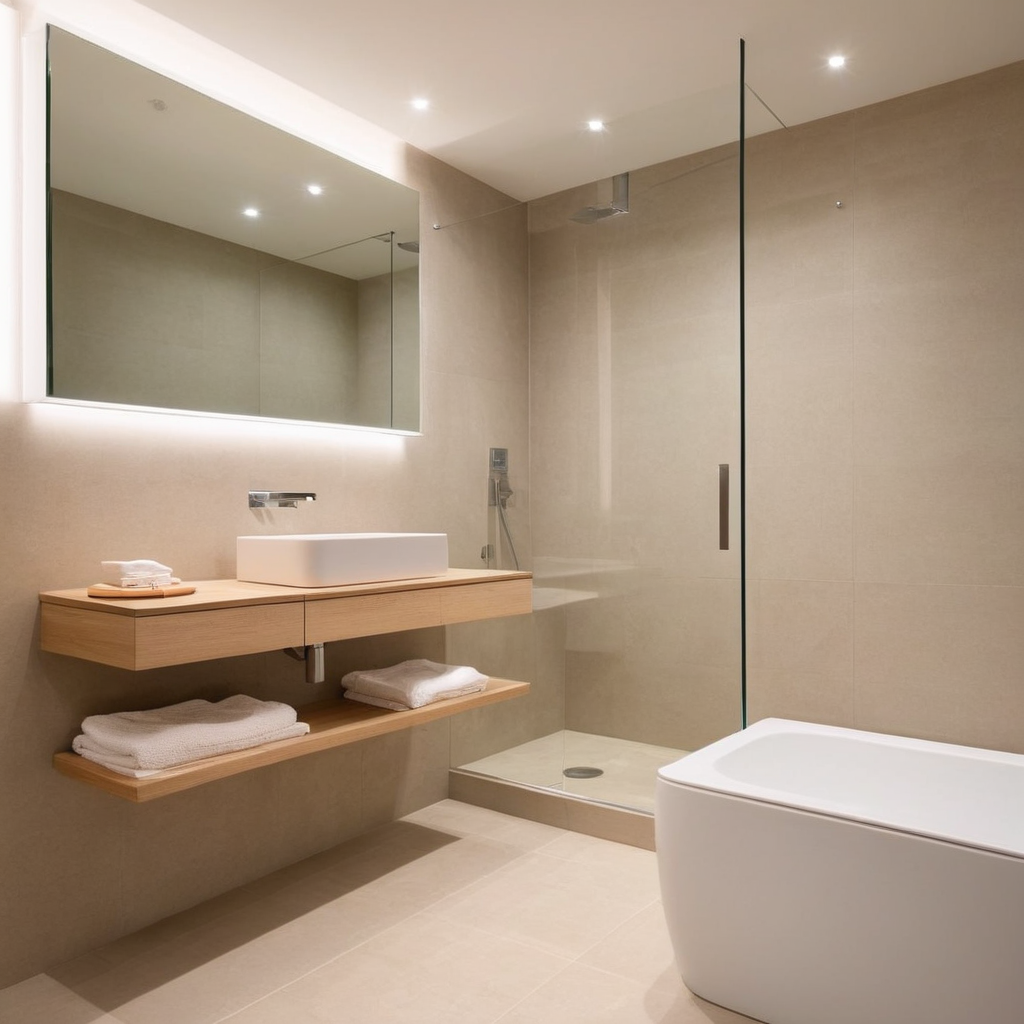}};

  \makerectanglecorners{\imagename}{0.2373}{0.6504}{0.4102}{0.7949};
  \drawbbox{firstcorner}{secondcorner}{\imagesource}{\imagewidth};

  \makeandplacecrop{\imagename}{-0.45}{0.7}{\imagesource}{\imagewidth}{2};
  \node[below =0cm of cropbox, align=center] (caption) {reflection};

  \makebboxline{bbox.west}{cropbox.east}

  \makerectanglecorners{\imagename}{ 0.5518}{0.7529}{0.6523}{0.8467};
  \drawbbox{firstcorner}{secondcorner}{\imagesource}{\imagewidth};

  \makeandplacecrop{\imagename}{1.1}{0.7}{\imagesource}{\imagewidth}{3};
  \node[below =0cm of cropbox, align=center] (caption) {irregular shape};

  \makebboxline{bbox.east}{cropbox.west}

  \makerectanglecorners{\imagename}{0.4756}{0.4932}{0.5117}{0.5674};
  \drawbbox{firstcorner}{secondcorner}{\imagesource}{\imagewidth};

  \makeandplacecrop{\imagename}{1.4}{0.3}{\imagesource}{\imagewidth}{4};
  \node[below =0cm of cropbox, align=center] (caption) {lack of detail};

  \makebboxline{bbox.east}{cropbox.west}

  \makerectanglecorners{\imagename}{0.37}{0.1758}{0.4717}{0.4014};
  \drawbbox{firstcorner}{secondcorner}{\imagesource}{\imagewidth};

  \makeandplacecrop{\imagename}{1.1}{0.1}{\imagesource}{\imagewidth}{1.5};
  \node[below =0cm of cropbox, align=center] (caption) {shadow};

  \makebboxline{bbox.east}{cropbox.west}

  \makerectanglecorners{\imagename}{0.2803}{0.3408}{0.3184}{0.4307};
  \drawbbox{firstcorner}{secondcorner}{\imagesource}{\imagewidth};

  \makeandplacecrop{\imagename}{-0.25}{0.1}{\imagesource}{\imagewidth}{3};
  \node[below =0cm of cropbox, align=center] (caption) {semantic};

  \makebboxline{bbox.west}{cropbox.east}

  \makerectanglecorners{\imagename}{0.2344}{0.4805}{0.3252}{0.5244};
  \drawbbox{firstcorner}{secondcorner}{\imagesource}{\imagewidth};

  \makeandplacecrop{\imagename}{-0.55}{0.45}{\imagesource}{\imagewidth}{3};
  \node[below =0cm of cropbox, align=center] (caption) {misshapen};

  \makebboxline{bbox.west}{cropbox.east}

\end{tikzpicture}

\caption{Synthetic image of a bath room with annotated artifacts. The reflection in the mirror seems to show part of the wall that is next to and behind the mirror which is physically impossible. }
\label{fig:indoors_example}
\end{figure}

%% file: figures/indoors_vanishing_point_2.tex
\begin{figure}[ht]
\centering

\begin{tikzpicture}[anchor=south west]
\def\imagewidth{0.45\textwidth}
\def\imagesource{images/indoors/real_photos/steven-ungermann-CVTmLMv5oG4-unsplash.jpg}
\node[name=img, inner sep=0]{\includegraphics[width=\imagewidth]{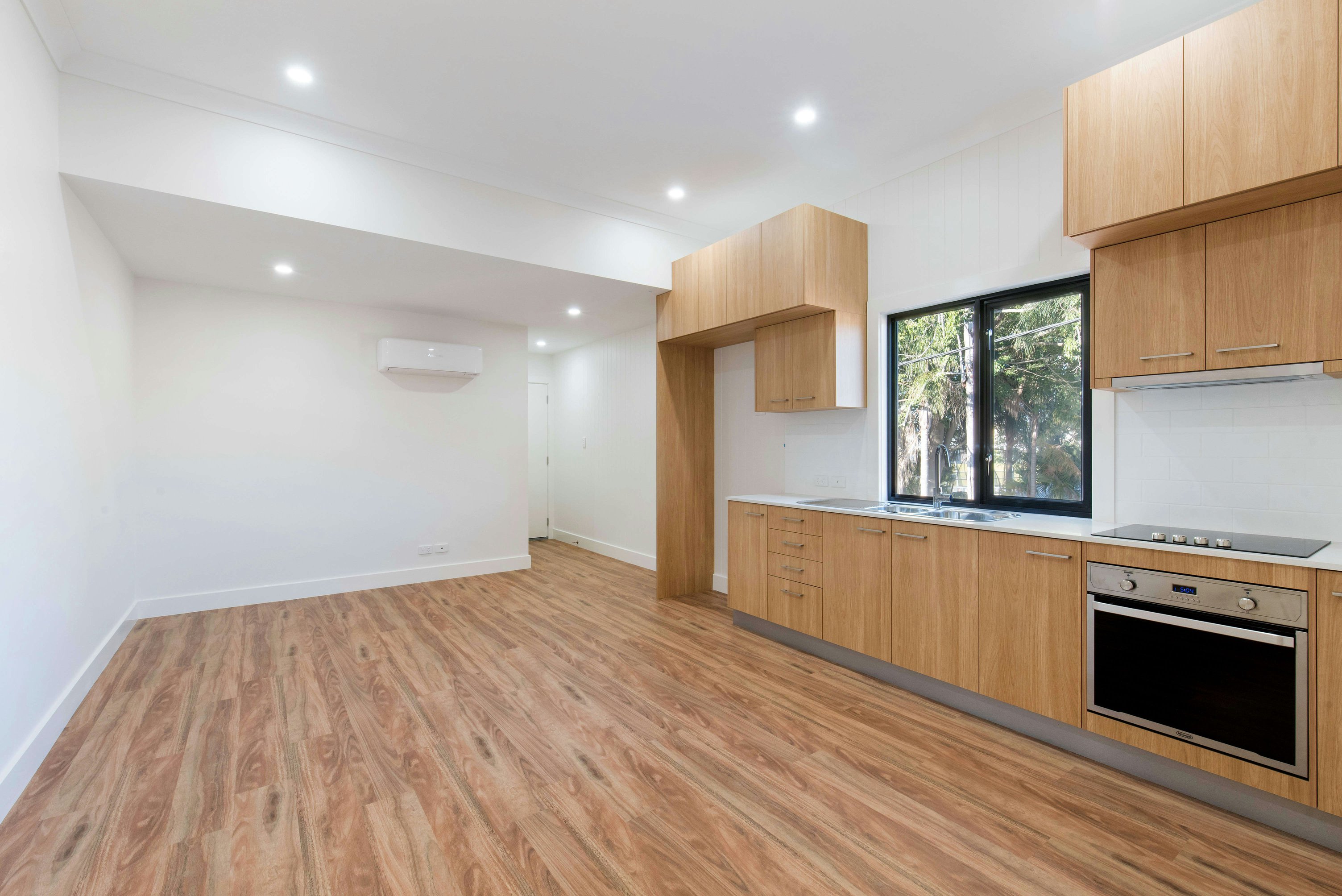}};

\makerectanglecorners{img}
{1.0}{0.087}{0.16}{0.519};
\draw[line width=5mm, \bboxcolor, \bboxbordersize, anchor=img.south west] (firstcorner) -- (secondcorner);

\makerectanglecorners{img}
{1.0}{0.36}{0.16}{0.51};
\draw[line width=5mm, \bboxcolor, \bboxbordersize, anchor=img.south west] (firstcorner) -- (secondcorner);

\makerectanglecorners{img}
{1.0}{0.58}{0.16}{0.505};
\draw[line width=5mm, \bboxcolor, \bboxbordersize, anchor=img.south west] (firstcorner) -- (secondcorner);

\makerectanglecorners{img}
{1.0}{0.7841}{0.16}{0.5};
\draw[line width=5mm, \bboxcolor, \bboxbordersize, anchor=img.south west] (firstcorner) -- (secondcorner);

\makerectanglecorners{img}
{1.0}{0.813}{0.16}{0.4978};
\draw[line width=5mm, \bboxcolor, \bboxbordersize, anchor=img.south west] (firstcorner) -- (secondcorner);

\makerectanglecorners{img}
{0.937}{0.99}{0.16}{0.49};
\draw[line width=5mm, \bboxcolor, \bboxbordersize, anchor=img.south west] (firstcorner) -- (secondcorner);
\end{tikzpicture}
\quad
\begin{tikzpicture}[anchor=south west]
\def\imagewidth{0.3\textwidth}
\def\imagesource{images/indoors/apartments/vanishing_point_example_3.png}
\node[name=img, inner sep=0]{\includegraphics[width=\imagewidth]{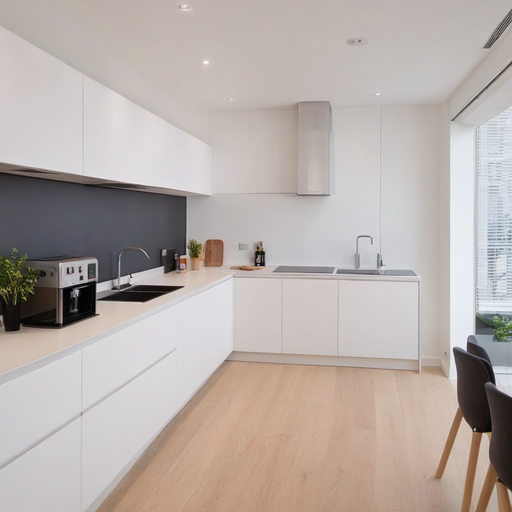}};

\makerectanglecorners{img}
{0.0}{0.9512}{0.712}{0.54};
\draw[line width=5mm, \bboxcolor, \bboxbordersize, anchor=img.south west] (firstcorner) -- (secondcorner);

\makerectanglecorners{img}
{0.0}{0.6836}{0.712}{0.57};
\draw[line width=5mm, \bboxcolor, \bboxbordersize, anchor=img.south west] (firstcorner) -- (secondcorner);

\makerectanglecorners{img}
{0.0}{0.2655}{0.8535}{0.641};
\draw[line width=5mm, \bboxcolor, \bboxbordersize, anchor=img.south west] (firstcorner) -- (secondcorner);

\makerectanglecorners{img}
{0.1621}{0.001}{0.8535}{0.7375};
\draw[line width=5mm, \bboxcolor, \bboxbordersize, anchor=img.south west] (firstcorner) -- (secondcorner);

\makerectanglecorners{img}
{0.0}{0.6675}{0.75}{0.56};
\draw[line width=5mm, \bboxcolor, \bboxbordersize, anchor=img.south west] (firstcorner) -- (secondcorner);

\makerectanglecorners{img}
{0.1865}{0.446}{0.7451}{0.5967};
\draw[line width=5mm, \bboxcolor, \bboxbordersize, anchor=img.south west] (firstcorner) -- (secondcorner);

\end{tikzpicture}

\caption{Comparison of vanishing points in real (left) (\href{https://unsplash.com/de/fotos/brauner-parkettboden-CVTmLMv5oG4}{Source: Steven Ungermann/Unsplash}) and generated (right) photographs. Parallel lines on the same plane all converge to the same vanishing point, as can be seen in the real photograph. However, in the generated image, the parallel lines do not converge to a single vanishing point. The deviation seems too large to be explained by a camera-related cause such as lens distortion.}
\label{fig:indoor_vanishing_points_2}

\end{figure}

%% file: figures/indoors_apartments.tex
\begin{figure}[ht]
\centering
    \begin{tikzpicture}[anchor=south west]
  \def\imagewidth{0.22\textwidth}
  \def\imagesource{images/indoors/apartments/sdxl_juggernaut_xl_v8_1.png}
  \node[name=img, inner sep=0]{\includegraphics[width=\imagewidth]{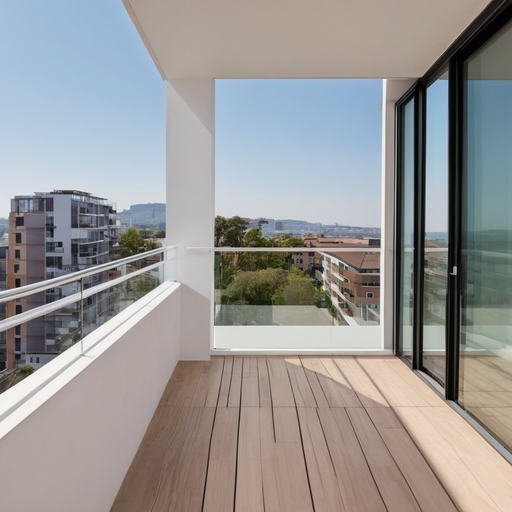}};
  \end{tikzpicture}
  \quad
  \begin{tikzpicture}[anchor=south west]
  \def\imagewidth{0.22\textwidth}
  \def\imagesource{images/indoors/apartments/sdxl_juggernaut_xl_v8_3.png}
  \node[name=img, inner sep=0]{\includegraphics[width=\imagewidth]{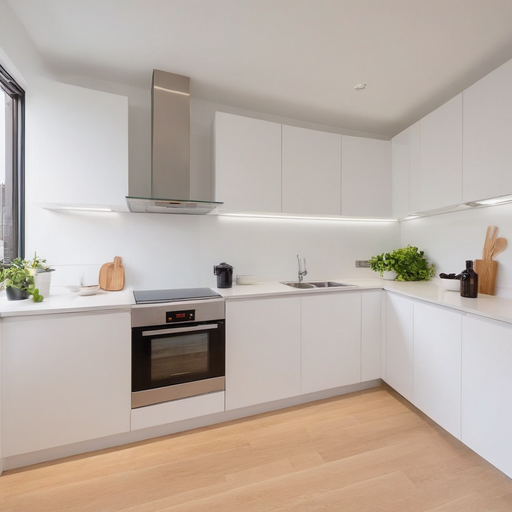}};
  \end{tikzpicture}
  \quad
  \begin{tikzpicture}[anchor=south west]
  \def\imagewidth{0.22\textwidth}
  \def\imagesource{images/indoors/apartments/sdxl_juggernaut_xl_v8_4.png}
  \node[name=img, inner sep=0]{\includegraphics[width=\imagewidth]{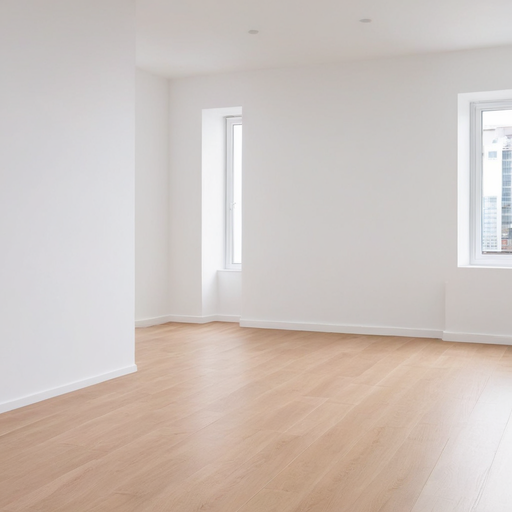}};
  \end{tikzpicture}
  \quad
  \begin{tikzpicture}[anchor=south west]
  \def\imagewidth{0.22\textwidth}
  \def\imagesource{images/indoors/apartments/sdxl_juggernaut_xl_v8_5.png}
  \node[name=img, inner sep=0]{\includegraphics[width=\imagewidth]{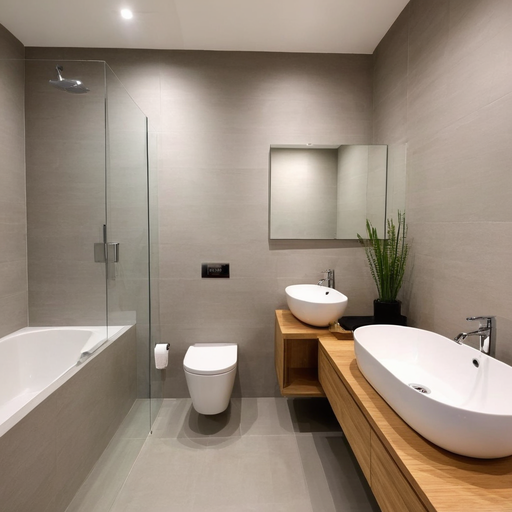}};
  \end{tikzpicture}
  \caption{Images of apartment rooms showing various artifacts, mainly related to geometry, as well as semantics.}
  \label{fig:indoor_apartments}
\end{figure}

%% file: figures/indoors_underground.tex
\begin{figure}[h]
\centering
    \begin{tikzpicture}[anchor=south west]
  \def\imagewidth{0.22\textwidth}
  \def\imagesource{images/indoors/tunnels/sdxl_juggernaut_xl_v8_1.png}
  \node[name=img, inner sep=0]{\includegraphics[width=\imagewidth]{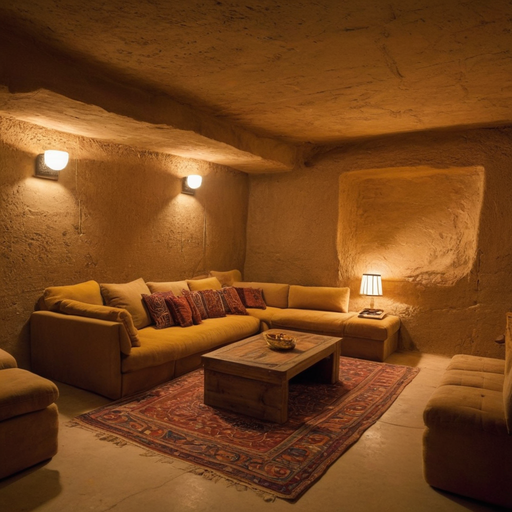}};
  \end{tikzpicture}
  \quad
  \begin{tikzpicture}[anchor=south west]
  \def\imagewidth{0.22\textwidth}
  \def\imagesource{images/indoors/tunnels/sdxl_juggernaut_xl_v8_2.png}
  \node[name=img, inner sep=0]{\includegraphics[width=\imagewidth]{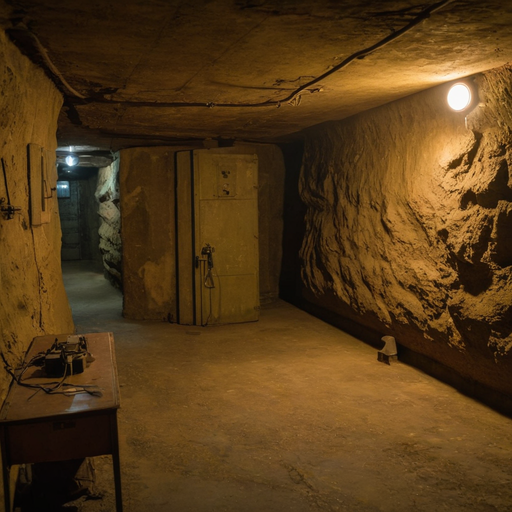}};
  \end{tikzpicture}
  \quad
  \begin{tikzpicture}[anchor=south west]
  \def\imagewidth{0.22\textwidth}
  \def\imagesource{images/indoors/tunnels/sdxl_juggernaut_xl_v8_3.png}
  \node[name=img, inner sep=0]{\includegraphics[width=\imagewidth]{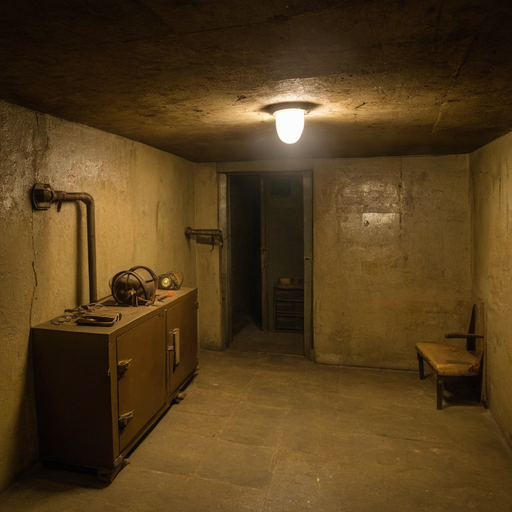}};
  \end{tikzpicture}
  \quad
  \begin{tikzpicture}[anchor=south west]
  \def\imagewidth{0.22\textwidth}
  \def\imagesource{images/indoors/tunnels/sdxl_juggernaut_xl_v8_4.png}
  \node[name=img, inner sep=0]{\includegraphics[width=\imagewidth]{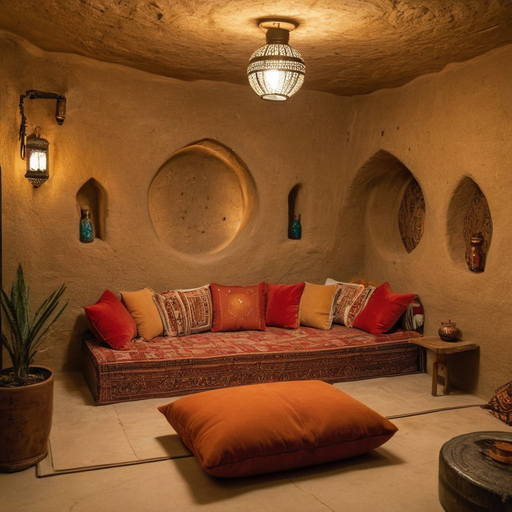}};
  \end{tikzpicture}
  \caption{Images of underground bunker or tunnel rooms with various artifacts related to lighting, as well as irregular cushions.}
  \label{fig:indoor_bunkers}
\end{figure}

%% file: figures/outdoor_intro.tex
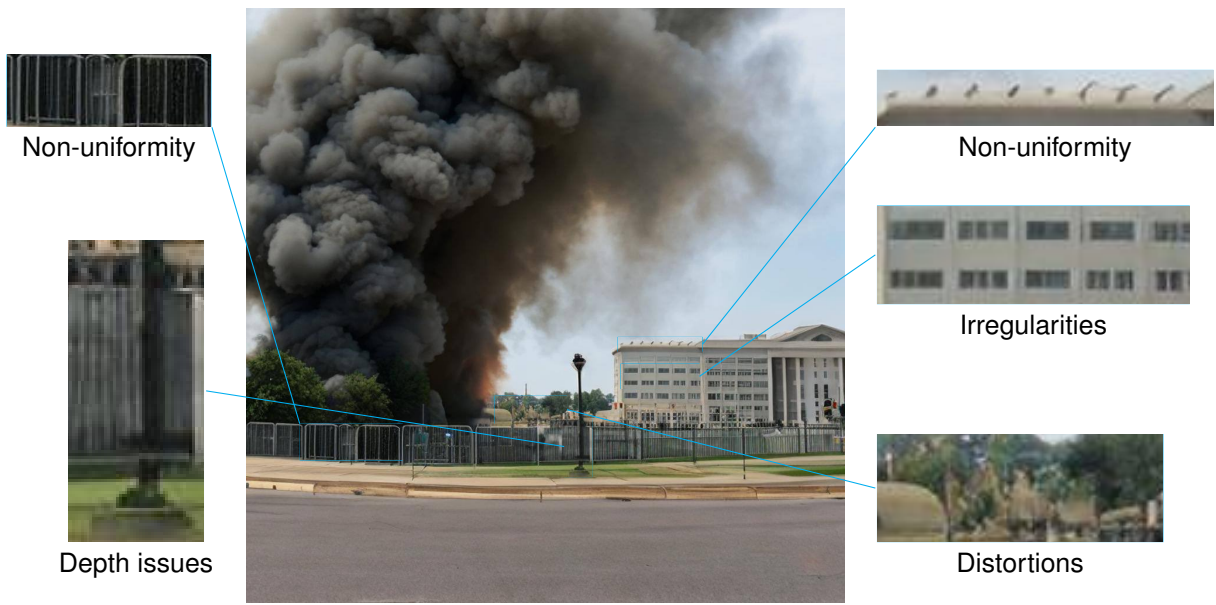
\begin{figure}[h]
    \centering
    \begin{tikzpicture}[anchor=south west]
        \def\imagename{img2}
        \def\imagewidth{0.5\textwidth}
        \def\imagesource{images/outdoors/pentagon_burning2.jpeg}
        \node[name=\imagename, inner sep=0, anchor=south west]{\includegraphics[width=\imagewidth]{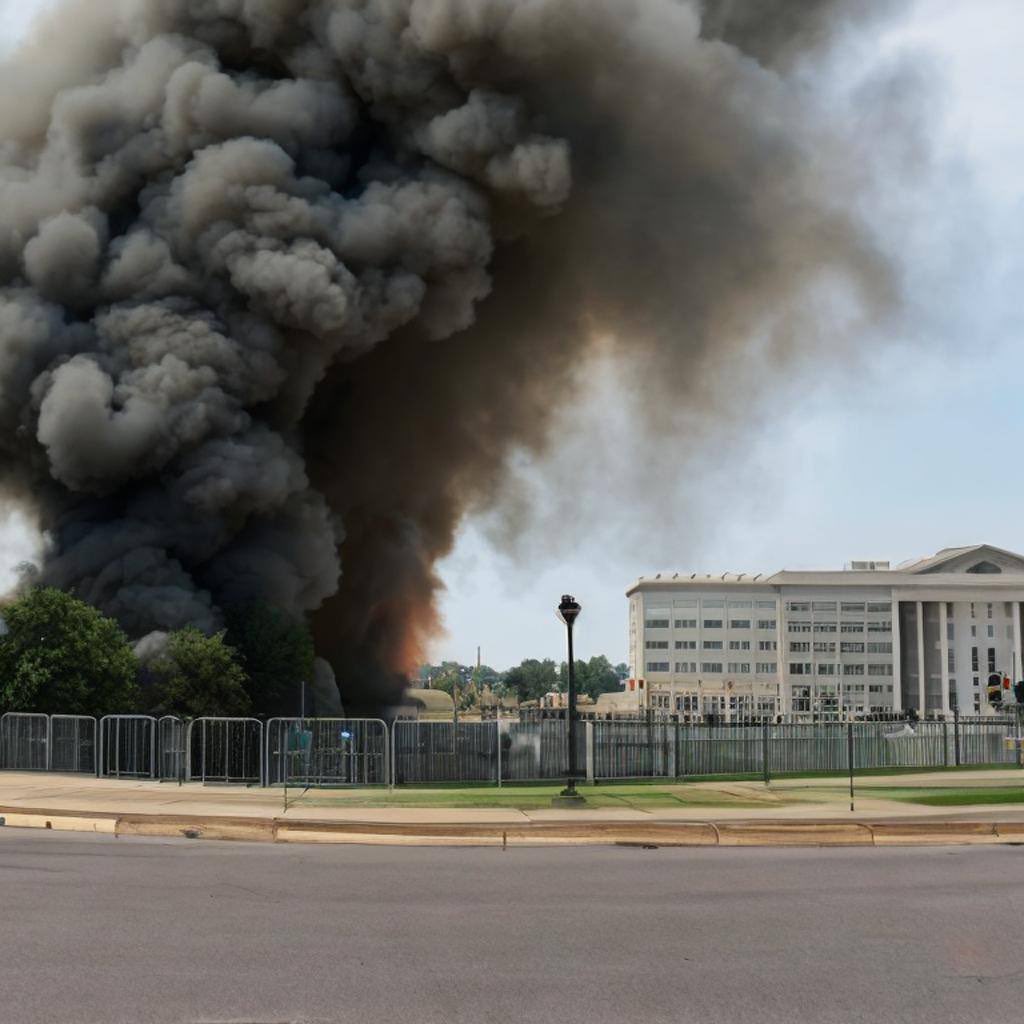}};

        \makerectanglecorners{\imagename}{0.09}{0.24}{0.26}{0.30};
        \drawbbox{firstcorner}{secondcorner}{\imagesource}{\imagewidth};

        \makeandplacecrop{\imagename}{-0.4}{0.8}{\imagesource}{\imagewidth}{2};

        \node[below =0cm of cropbox, align=center] (caption) {Non-uniformity};

        \makebboxline{bbox.north west}{cropbox.south east}

        \makerectanglecorners{\imagename}{0.625}{0.360}{0.755}{0.401};
        \drawbbox{firstcorner}{secondcorner}{\imagesource}{\imagewidth};

        \makeandplacecrop{\imagename}{1.05}{0.5}{\imagesource}{\imagewidth}{4};

        \node[below =0cm of cropbox, align=center] (caption) {Irregularities};

        \makebboxline{bbox.east}{cropbox.west}

        \makerectanglecorners{\imagename}{0.6201}{0.423}{0.760}{0.446};
        \drawbbox{firstcorner}{secondcorner}{\imagesource}{\imagewidth};

        \makeandplacecrop{\imagename}{1.05}{0.8}{\imagesource}{\imagewidth}{4};

        \node[below =0cm of cropbox, align=center] (caption) {Non-uniformity};

        \makebboxline{bbox.east}{cropbox.south west}

        \makerectanglecorners{\imagename}{0.531}{0.212}{0.576}{0.313};
        \drawbbox{firstcorner}{secondcorner}{\imagesource}{\imagewidth};

        \makeandplacecrop{\imagename}{-0.3}{0.1}{\imagesource}{\imagewidth}{5};

        \node[below =0cm of cropbox, align=center] (caption) {Depth issues};

        \makebboxline{bbox.west}{cropbox.east}

        \makerectanglecorners{\imagename}{0.415}{0.302}{0.534}{0.347};
        \drawbbox{firstcorner}{secondcorner}{\imagesource}{\imagewidth};

        \makeandplacecrop{\imagename}{1.05}{0.1}{\imagesource}{\imagewidth}{4};

        \node[below =0cm of cropbox, align=center] (caption) {Distortions};

        \makebboxline{bbox.east}{cropbox.west}

      \end{tikzpicture}
      \caption{Synthetic photography purportedly showing an explosion / fire near the Pentagon. There are multiple artifacts related to lack of uniformity and regularity when such is expected, as well as texture distortions and geometrical inconsistencies ( \href{https://x.com/RoiLopezRivas/status/1660649454012080129}{Source: X/Twitter})}
      \label{fig:outdoor_intro}
\end{figure}

%% file: figures/outdoor_vanishing_point.tex
\begin{figure}[ht]
\centering

\begin{tikzpicture}[anchor=south west]
\def\imagewidth{0.5\textwidth}
\def\imagesource{images/outdoors/real_photos/claudio-schwarz-DpeXitxtix8-unsplash.jpg}
\node[name=img, inner sep=0]{\includegraphics[width=\imagewidth]{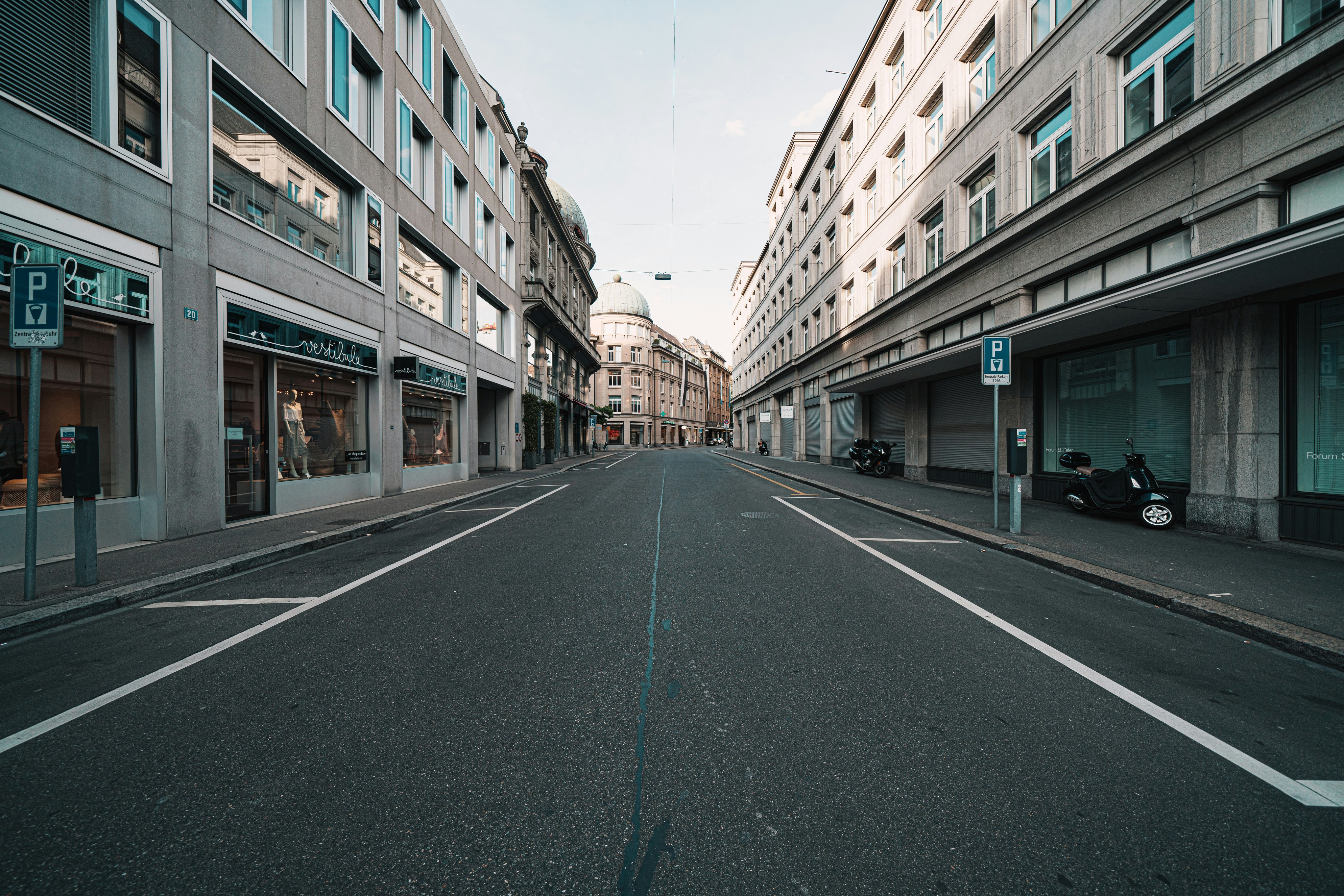}};

\makerectanglecorners{img}
{0.975}{ 0.0982}{0.5}{0.5098};
\draw[line width=5mm, \bboxcolor, \bboxbordersize, anchor=img.south west] (firstcorner) -- (secondcorner);

\makerectanglecorners{img}
{1}{0.7702}{0.5}{0.5098};
\draw[line width=5mm, \bboxcolor, \bboxbordersize, anchor=img.south west] (firstcorner) -- (secondcorner);

\makerectanglecorners{img}
{0.8179}{0.9957}{0.5}{0.5098};
\draw[line width=5mm, \bboxcolor, \bboxbordersize, anchor=img.south west] (firstcorner) -- (secondcorner);

\makerectanglecorners{img}
{0.0}{0.1565}{0.5}{0.5098};
\draw[line width=5mm, \bboxcolor, \bboxbordersize, anchor=img.south west] (firstcorner) -- (secondcorner);

\makerectanglecorners{img}
{0.0}{0.7911}{0.5}{0.5098};
\draw[line width=5mm, \bboxcolor, \bboxbordersize, anchor=img.south west] (firstcorner) -- (secondcorner);

\makerectanglecorners{img}
{0.3248}{1.0004}{0.5}{0.5098};
\draw[line width=5mm, \bboxcolor, \bboxbordersize, anchor=img.south west] (firstcorner) -- (secondcorner);
\end{tikzpicture}
\begin{tikzpicture}[anchor=south west]
\def\imagewidth{0.35\textwidth}
\def\imagesource{images/outdoors/city_streets/sdxl_juggernaut_xl_v8_123VR.png}
\node[name=img, inner sep=0]{\includegraphics[width=\imagewidth]{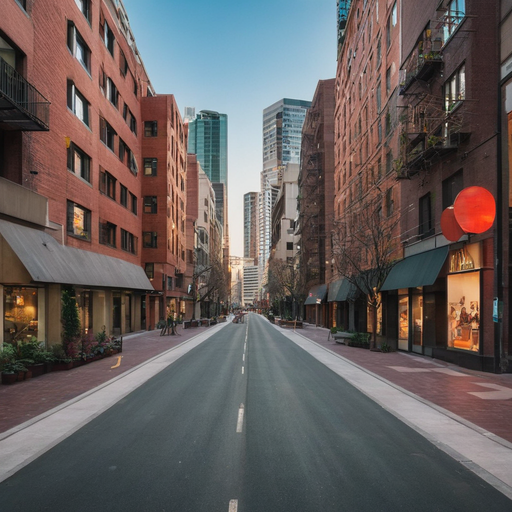}};

\makerectanglecorners{img}
{0.9961}{0.0205}{0.4502}{0.415};
\draw[line width=5mm, \bboxcolor, \bboxbordersize, anchor=img.south west] (firstcorner) -- (secondcorner);

\makerectanglecorners{img}
{0.0029}{ 0.0557}{0.5264}{0.4238};
\draw[line width=5mm, \bboxcolor, \bboxbordersize, anchor=img.south west] (firstcorner) -- (secondcorner);

\makerectanglecorners{img}
{0.1943}{0.626}{ 0.541}{0.4082};
\draw[line width=5mm, \bboxcolor, \bboxbordersize, anchor=img.south west] (firstcorner) -- (secondcorner);

\makerectanglecorners{img}
{0.916}{0.5293}{0.415}{0.3936};
\draw[line width=5mm, \bboxcolor, \bboxbordersize, anchor=img.south west] (firstcorner) -- (secondcorner);

\makerectanglecorners{img}
{0.3359}{0.8164}{0.5332}{0.2939};
\draw[line width=5mm, \bboxcolor, \bboxbordersize, anchor=img.south west] (firstcorner) -- (secondcorner);

\end{tikzpicture}

\caption{Comparison of vanishing points in real (left) (\href{https://unsplash.com/de/fotos/leere-strasse-zwischen-hochhausern-tagsuber-DpeXitxtix8}{Source: Claudio Schwarz/Unsplash}) and generated (right) photographs. Parallel lines on the same plane all converge to the same vanishing point, as can be seen in the real photograph. However, in the generated image, the parallel lines do not converge to a single vanishing point}
\label{fig:outdoor_vanishing_points}

\end{figure}
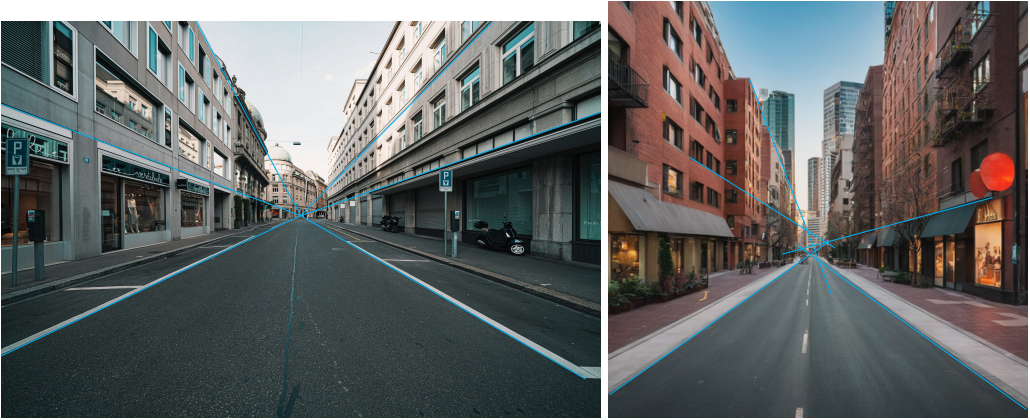

%% file: figures/outdoor_war_zone.tex
\begin{figure}[ht]
\centering
    \begin{tikzpicture}[anchor=south west]
  \def\imagewidth{0.3\textwidth}
  \def\imagesource{images/outdoors/war_scenes/sdxl_juggernaut_xl_v8_cctv_1.png}
  \node[name=img, inner sep=0]{\includegraphics[width=\imagewidth]{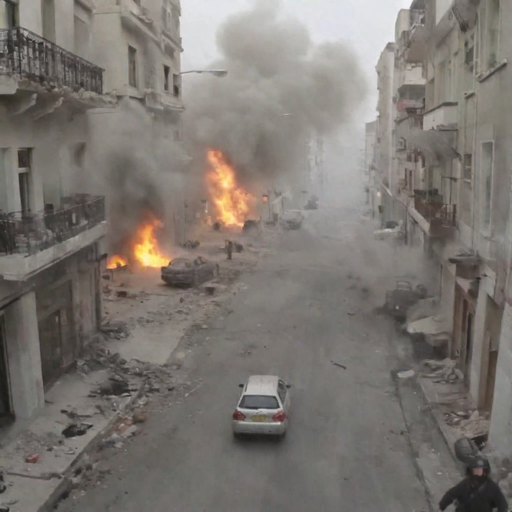}};
  \end{tikzpicture}
  \quad
  \begin{tikzpicture}[anchor=south west]
  \def\imagewidth{0.3\textwidth}
\def\imagesource{images/outdoors/war_scenes/sdxl_juggernaut_xl_v8_cctv_2.png}
  \node[name=img, inner sep=0]{\includegraphics[width=\imagewidth]{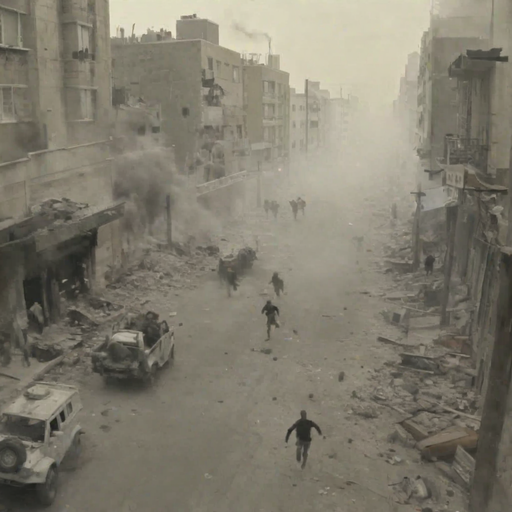}};
  \end{tikzpicture}
\quad
  \begin{tikzpicture}[anchor=south west]
  \def\imagewidth{0.3\textwidth}
\def\imagesource{images/outdoors/war_scenes/sdxl_juggernaut_xl_v8_cctv_4.png}
  \node[name=img, inner sep=0]{\includegraphics[width=\imagewidth]{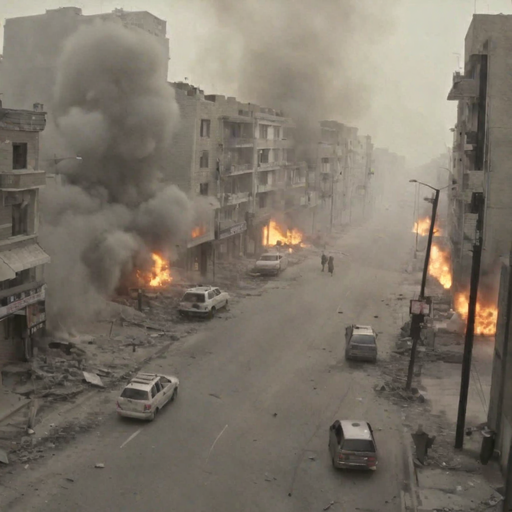}};
  \end{tikzpicture}
  \caption{Synthetic images of war scenes in the streets of a city recorded by cctv. From left to right: The first image depicts a person at the lower right corner which seems unexpectedly large when compared to the car. The second image exhibits low detail of the people in the streets, while the third image exhibits asymmetries in the vehicles.}
   \label{fig:outdoor_war_zone_city}
\end{figure}

%% file: figures/outdoor_water_park.tex
\begin{figure}[ht]
\centering
    \begin{tikzpicture}[anchor=south west]
  \def\imagewidth{0.3\textwidth}
  \def\imagesource{images/outdoors/fake_vacation/social_media_1.jpg}
  \node[name=img, inner sep=0]{\includegraphics[width=\imagewidth]{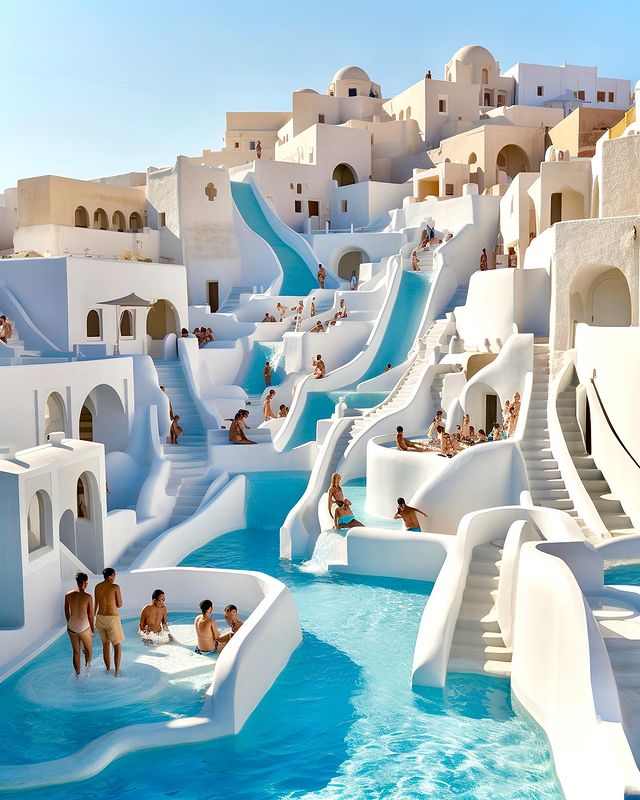}};
  \end{tikzpicture}
  \quad
      \begin{tikzpicture}[anchor=south west]
  \def\imagewidth{0.33\textwidth}
  \def\imagesource{images/outdoors/fake_vacation/social_media_2.jpg}
  \node[name=img, inner sep=0]{\includegraphics[width=\imagewidth]{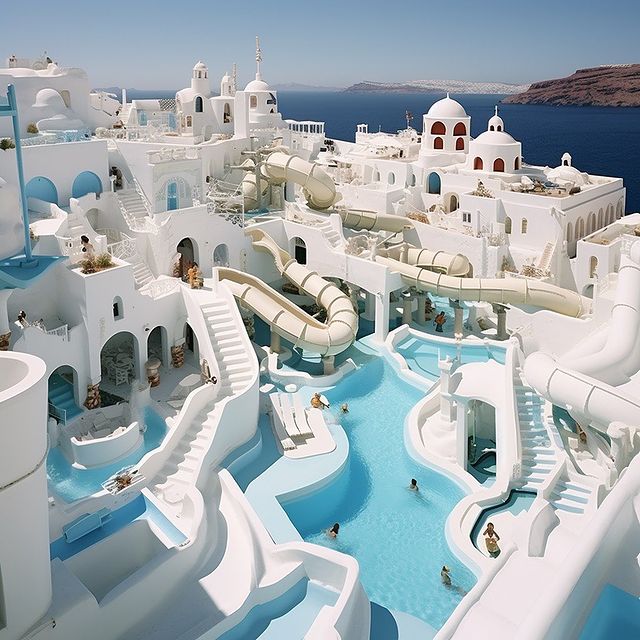}};
  \end{tikzpicture}
  \quad
      \begin{tikzpicture}[anchor=south west]
  \def\imagewidth{0.3\textwidth}
  \def\imagesource{images/outdoors/fake_vacation/social_media_3.jpg}
  \node[name=img, inner sep=0]{\includegraphics[width=\imagewidth]{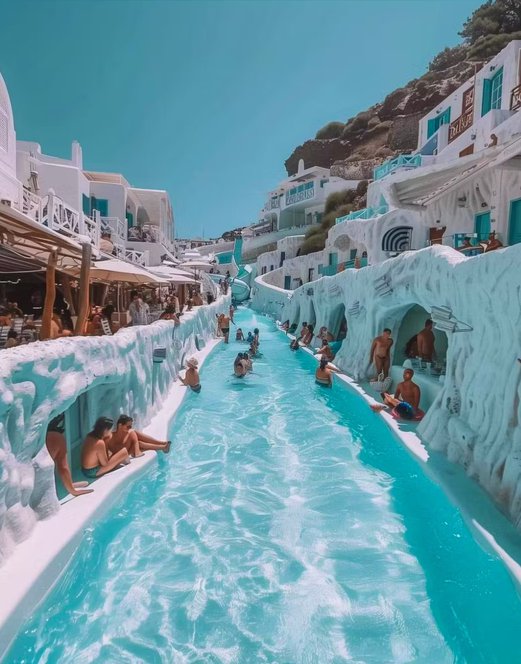}};
  \end{tikzpicture}
  \caption{Generated images of a vacation resort with a water park whose building style resembles that of Santorini and which circulated on social media (Source: \cite{riebeling_santorini_2024}).}
  \label{fig:outdoor_water_park}
\end{figure}

%% file: figures/outdoor_fake_vacation.tex
\begin{figure}[ht]
\centering
    \begin{tikzpicture}[anchor=south west]
  \def\imagewidth{0.3\textwidth}
  \def\imagesource{images/outdoors/fake_vacation/sdxl_juggernaut_xl_v8_1.png}
  \node[name=img, inner sep=0]{\includegraphics[width=\imagewidth]{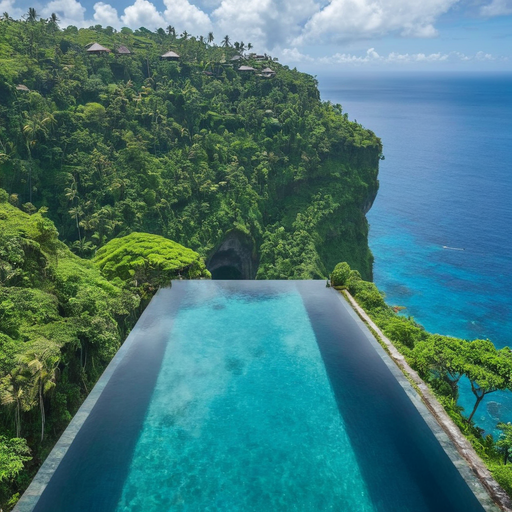}};
  \end{tikzpicture}
  \quad
  \begin{tikzpicture}[anchor=south west]
  \def\imagewidth{0.3\textwidth}
\def\imagesource{images/outdoors/fake_vacation/SD1.5_architectureExterior_v40_5.png}
  \node[name=img, inner sep=0]{\includegraphics[width=\imagewidth]{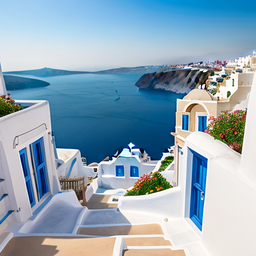}};
  \end{tikzpicture}
  \quad
  \begin{tikzpicture}[anchor=south west]
  \def\imagewidth{0.3\textwidth}
\def\imagesource{images/outdoors/fake_vacation/SD1.5_architectureExterior_v40_3.png}
  \node[name=img, inner sep=0]{\includegraphics[width=\imagewidth]{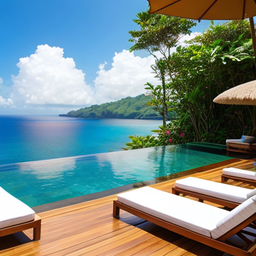}};
  \end{tikzpicture}
  \caption{Synthetic images from vacation spots and hotels. Overall, the coloring seems unnaturally intense. From left to right: The vegetation in the first image covers the end of the pool, which makes the pool appear distant and extremely large. The second image looks cartoonish and the lines of building parts and stairs are not straight. The pool boundary in the third image is not straight, the size of the (identical) deck chairs is inconsistent, and the reflections of the vegetation does not appear correct.}
  \label{fig:outdoor_fake_vacation}
\end{figure}

%% file: figures/objects_himars.tex
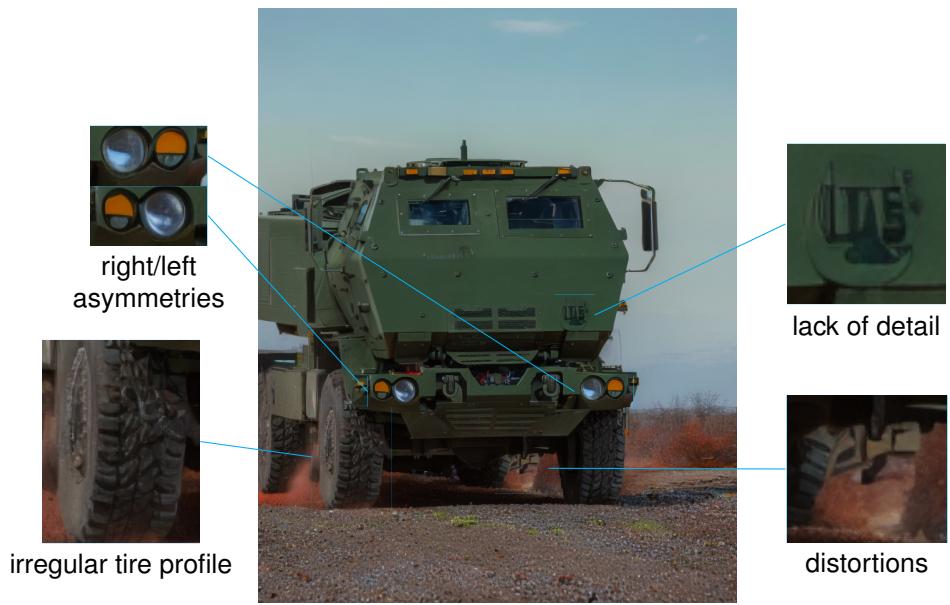
\begin{figure}[h]
  \center
  \begin{tikzpicture}[anchor=south west]
  \def\imagewidth{0.4\textwidth}
  \def\imagesource{images/items/himars_sample2.png}
  \def\imagename{img}

  \node[name=\imagename, inner sep=0]{\includegraphics[width=\imagewidth]{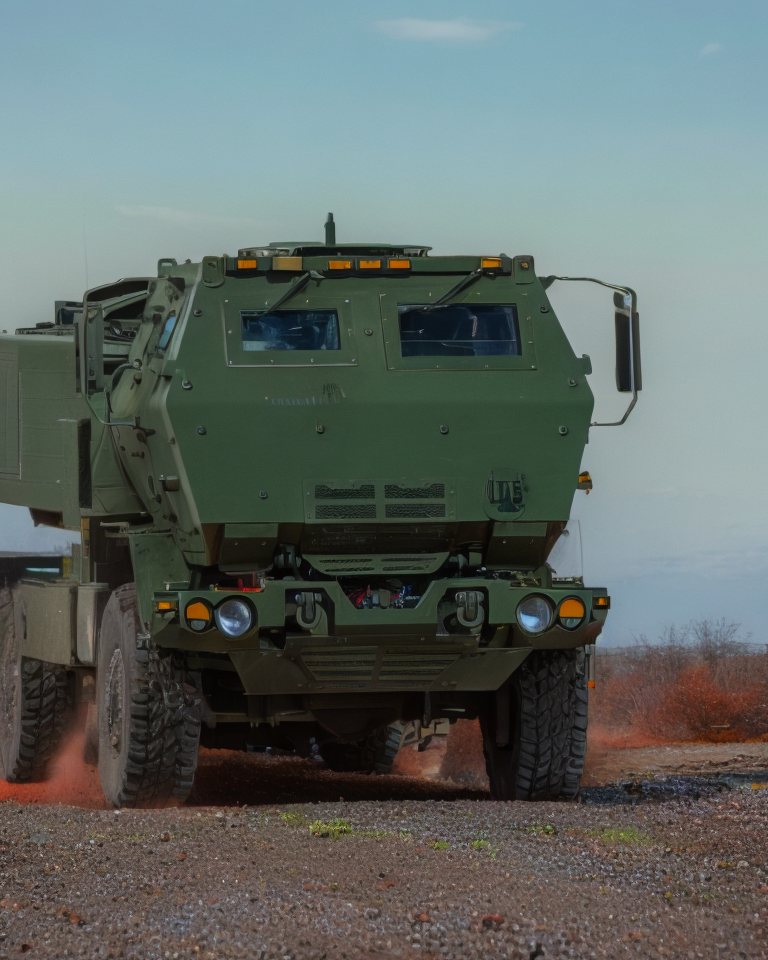}};

  \makerectanglecorners{\imagename}{0.113}{0.16}{0.276}{0.33};
  \drawbbox{firstcorner}{secondcorner}{\imagesource}{\imagewidth};

  \makeandplacecrop{\imagename}{-0.45}{0.1}{\imagesource}{\imagewidth}{2};
  \node[below =0cm of cropbox, align=center] (caption) {irregular tire profile};

  \makebboxline{bbox.west}{cropbox.east}

  \makerectanglecorners{\imagename}{0.229}{0.33}{0.35}{0.38};
  \drawbbox{firstcorner}{secondcorner}{\imagesource}{\imagewidth};

  \makeandplacecrop{\imagename}{-0.35}{0.6}{\imagesource}{\imagewidth}{2};
  \node[below =0cm of cropbox, align=center] (caption) {right/left\\asymmetries};

  \makebboxline{bbox.west}{cropbox.east}

  \makerectanglecorners{\imagename}{ 0.66}{0.33}{0.781}{0.38};
  \drawbbox{firstcorner}{secondcorner}{\imagesource}{\imagewidth};

  \makeandplacecrop{\imagename}{-0.35}{0.7}{\imagesource}{\imagewidth}{2};
  \node[below =0cm of cropbox, align=center] (caption) {};

  \makebboxline{bbox.west}{cropbox.east}

  \makerectanglecorners{\imagename}{0.493}{0.185}{0.604}{0.267};
  \drawbbox{firstcorner}{secondcorner}{\imagesource}{\imagewidth};

  \makeandplacecrop{\imagename}{1.1}{0.1}{\imagesource}{\imagewidth}{3};
  \node[below =0cm of cropbox, align=center] (caption) {distortions};

  \makebboxline{bbox.east}{cropbox.west}

  \makerectanglecorners{\imagename}{0.618}{0.45}{0.7}{0.517};
  \drawbbox{firstcorner}{secondcorner}{\imagesource}{\imagewidth};

  \makeandplacecrop{\imagename}{1.1}{0.5}{\imagesource}{\imagewidth}{4};
  \node[below =0cm of cropbox, align=center] (caption) {lack of detail};

  \makebboxline{bbox.east}{cropbox.west}

  \end{tikzpicture}

    \caption{Synthetic image of a HIMARS military vehicle \cite{mathys_synthetic_2024}.}
  \label{fig:objects_himars}
\end{figure}

%% file: figures/objects_vehicles.tex
\begin{figure}[ht]
\centering
    \begin{tikzpicture}[anchor=south west]
  \def\imagewidth{0.4\textwidth}
  \def\imagesource{images/items/vehicles_mjv5/r0a2ff882t.png}
  \node[name=img, inner sep=0]{\includegraphics[width=\imagewidth]{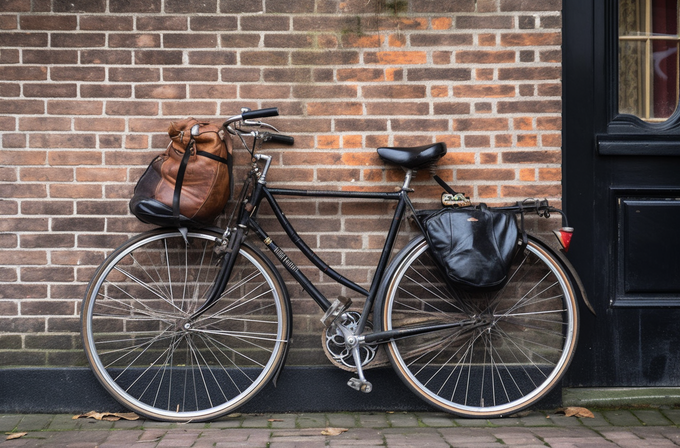}};
  \end{tikzpicture}
  \begin{tikzpicture}[anchor=south west]
  \def\imagewidth{0.4\textwidth}
\def\imagesource{images/items/vehicles_mjv5/r0fc39b66t.png}
  \node[name=img, inner sep=0]{\includegraphics[width=\imagewidth]{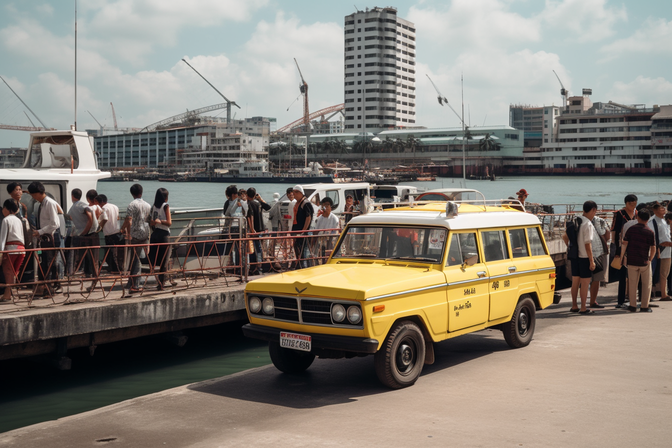}};
  \end{tikzpicture}
    \begin{tikzpicture}[anchor=south west]
  \def\imagewidth{0.4\textwidth}
\def\imagesource{images/items/vehicles_sdxl_sb/r1da851e4t.png}
  \node[name=img, inner sep=0]{\includegraphics[width=\imagewidth]{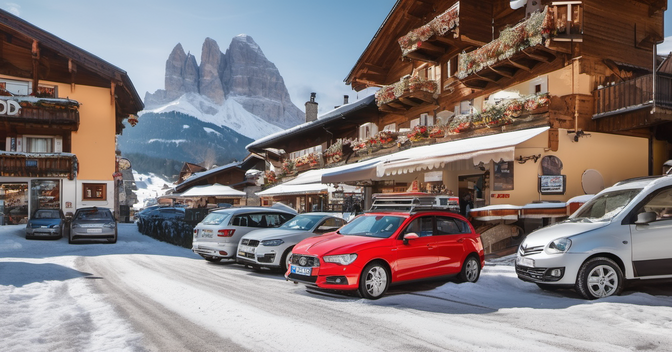}};
  \end{tikzpicture}
    \begin{tikzpicture}[anchor=south west]
  \def\imagewidth{0.4\textwidth}
\def\imagesource{images/items/vehicles_sdxl_sb/r1e79c6f9t.png}
  \node[name=img, inner sep=0]{\includegraphics[width=\imagewidth]{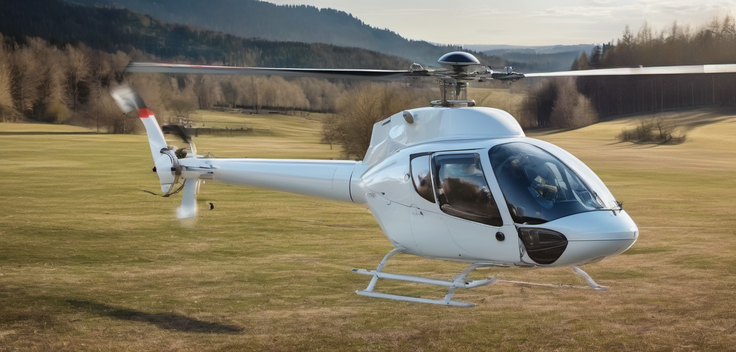}};
  \end{tikzpicture}

  \caption{Synthetic images of vehicles. Top left: Bicycle with non-uniform spokes, detached handle and inconsistent angle between pedals. Top right: Car with illegible plates, irregular tires, and misshaped roof rack. Bottom left: Cars with asymmetrical fronts, as well as local blur and lack of detail. Bottom right: Helicopter with misshaped tail rotor system and misshaped landing gear. All images from Synthbuster \cite{bammey_synthbuster_2023}, top-row generated with Midjourney v5, bottom-row with Stable Diffusion XL.}
  \label{fig:objects_vehicles}
\end{figure}

%% file: figures/objects_items.tex
\begin{figure}[h]
\centering
    \begin{tikzpicture}[anchor=south west]
  \def\imagewidth{0.22\textwidth}
  \def\imagesource{images/items/sdxl_juggernaut_xl_v8_1.png}
  \node[name=img, inner sep=0]{\includegraphics[width=\imagewidth]{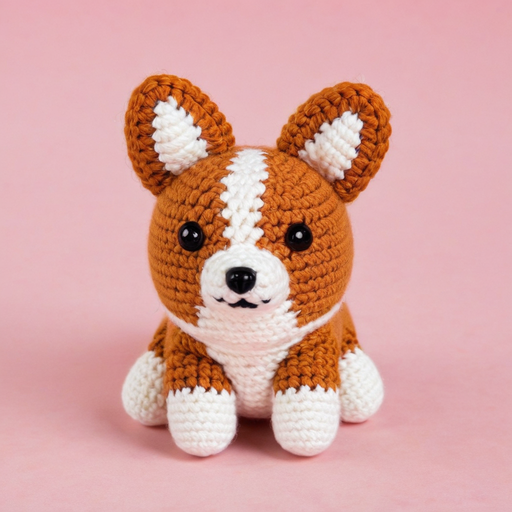}};
  \end{tikzpicture}
  \quad
  \begin{tikzpicture}[anchor=south west]
  \def\imagewidth{0.22\textwidth}
\def\imagesource{images/items/sdxl_juggernaut_xl_v8_2.png}
  \node[name=img, inner sep=0]{\includegraphics[width=\imagewidth]{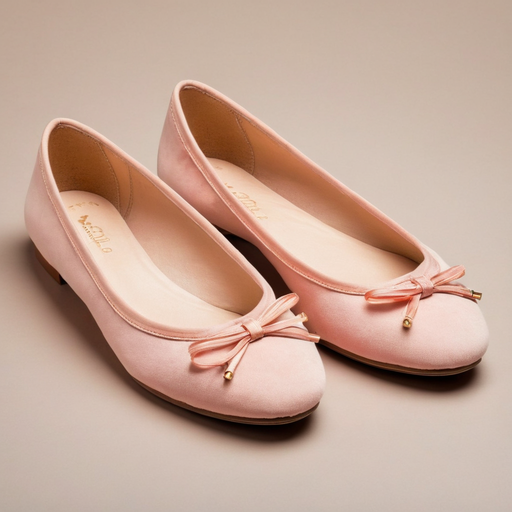}};
  \end{tikzpicture}
  \quad
  \begin{tikzpicture}[anchor=south west]
  \def\imagewidth{0.22\textwidth}
\def\imagesource{images/items/sdxl_juggernaut_xl_v8_3.png}
  \node[name=img, inner sep=0]{\includegraphics[width=\imagewidth]{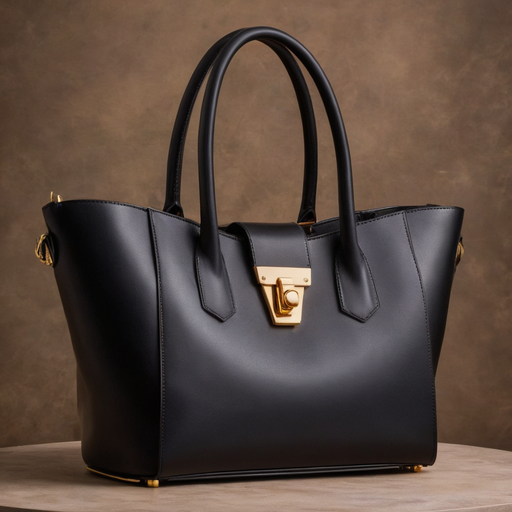}};
  \end{tikzpicture}
  \quad
  \begin{tikzpicture}[anchor=south west]
  \def\imagewidth{0.22\textwidth}
\def\imagesource{images/items/sdxl_juggernaut_xl_v8_4.png}
  \node[name=img, inner sep=0]{\includegraphics[width=\imagewidth]{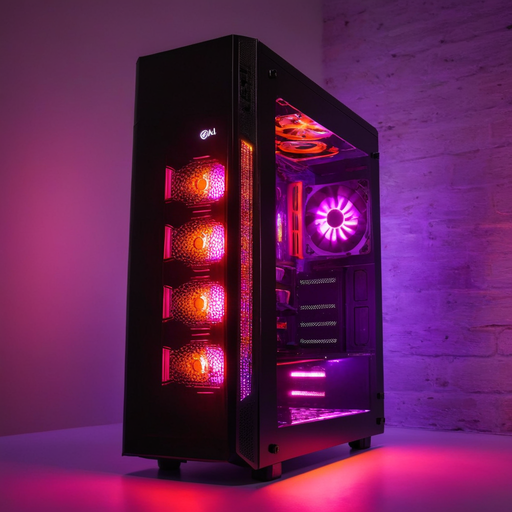}};
  \end{tikzpicture}
  \caption{Generated images of items. From left to right: A crochet puppy dog without clearly noticeable artifacts. A pair of shoes with illegible text, and slight asymmetries in the shape of the left and right shoe. A hand bag with irregular shapes in the buckle. A PC tower with semantic artifacts, such as missing power supply unit.}
  \label{fig:objects_items}
\end{figure}

%% file: figures/people_face_difficult.tex
\begin{figure}[h]
\centering
\begin{tikzpicture}[anchor=south west]
\def\imagename{img}
\def\imagewidth{0.5\textwidth}
\def\imagesource{images/humans/linkedin/SD15_realisticVision_linkedin_5.png}
\node[name=\imagename, inner sep=0]{\includegraphics[width=\imagewidth]{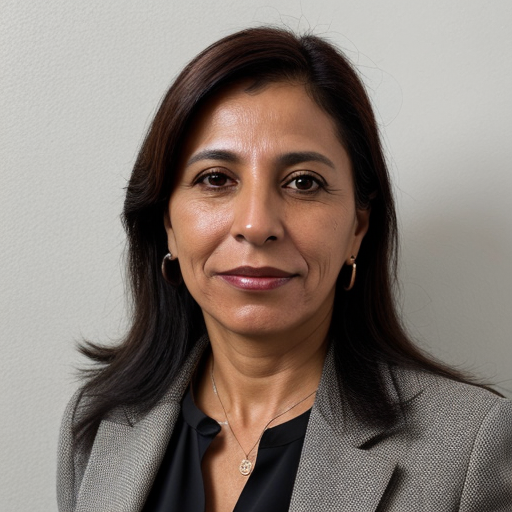}};

  \makerectanglecorners{\imagename}{0.45}{0.775}{0.55}{0.875};
  \drawbbox{firstcorner}{secondcorner}{\imagesource}{\imagewidth};

  \makeandplacecrop{\imagename}{1.1}{0.7}{\imagesource}{\imagewidth}{2.5};

  \node[below =0cm of cropbox, align=center] (caption) {Hair cut off?};

  \makebboxline{bbox.east}{cropbox.west}

  \makerectanglecorners{\imagename}{0.35}{0.6}{0.475}{0.725};
  \drawbbox{firstcorner}{secondcorner}{\imagesource}{\imagewidth};

  \makeandplacecrop{\imagename}{-0.6}{0.5}{\imagesource}{\imagewidth}{4};

  \node[below =0cm of cropbox, align=center] (caption) {Pupil shape?};

  \makebboxline{bbox.west}{cropbox.east}

  \makerectanglecorners{\imagename}{0.304}{0.439}{0.365}{0.517};
  \drawbbox{firstcorner}{secondcorner}{\imagesource}{\imagewidth};

  \makeandplacecrop{\imagename}{-0.4}{0.05}{\imagesource}{\imagewidth}{4};

  \node[below =0cm of cropbox, align=center] (caption) {Irregularity?};

  \makebboxline{bbox.west}{cropbox.east}

  \makerectanglecorners{\imagename}{0.369}{0.113}{0.554}{0.193};
  \drawbbox{firstcorner}{secondcorner}{\imagesource}{\imagewidth};

  \makeandplacecrop{\imagename}{1.1}{0.3}{\imagesource}{\imagewidth}{2};

  \node[below =0cm of cropbox, align=center] (caption) {Asymmetry?};

  \makebboxline{bbox.east}{cropbox.west}

\end{tikzpicture}
\caption{Synthetic portrait photography of a woman with small irregularities in the left eye, hair, clothing symmetry and earring form and attachment. }
\label{fig:people_face_difficult}
\end{figure}
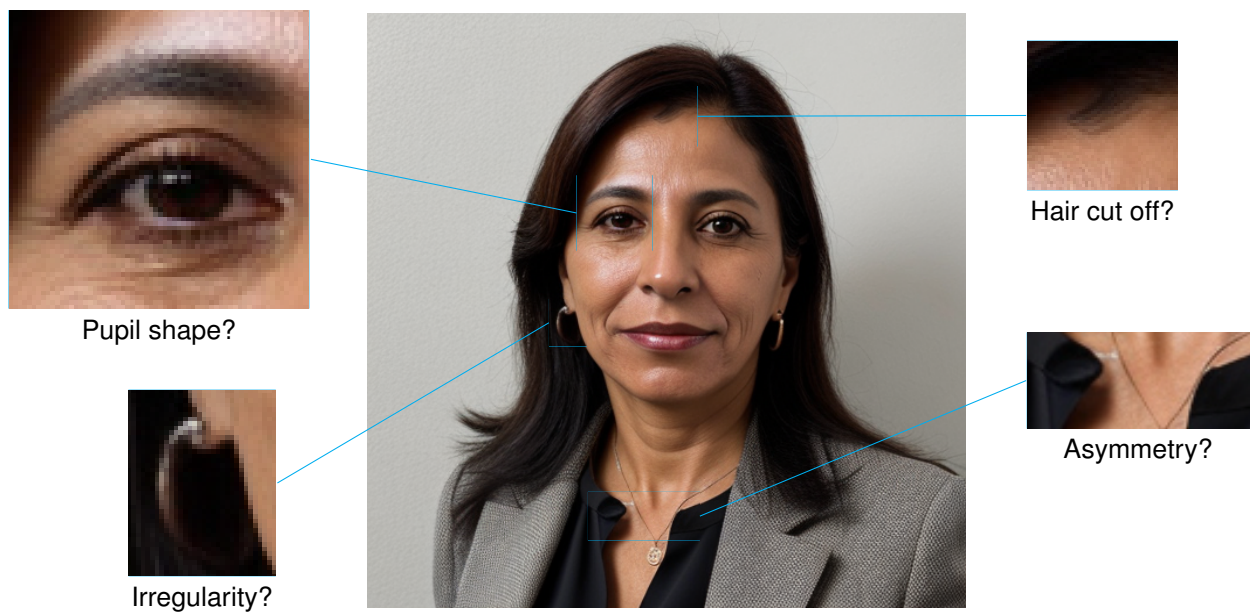

%% file: figures/difficult_items.tex
\begin{figure}[ht]
\centering
    \begin{subfigure}[b]{0.3\textwidth}
        \begin{tikzpicture}[anchor=south west]
            \def\imagewidth{\textwidth}
            \def\imagesource{images/items/sdxl_juggernaut_xl_v8_5.png}
            \node[name=img, inner sep=0]{\includegraphics[width=\imagewidth]{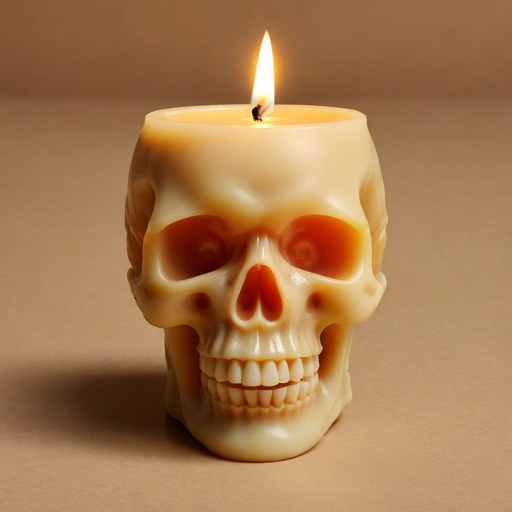}};
        \end{tikzpicture}
    \end{subfigure}
  \quad
    \begin{subfigure}[b]{0.3\textwidth}
        \begin{tikzpicture}[anchor=south west]
            \def\imagewidth{\textwidth}
            \def\imagesource{images/items/sdxl_juggernaut_xl_v8_6.png}
            \node[name=img, inner sep=0]{\includegraphics[width=\imagewidth]{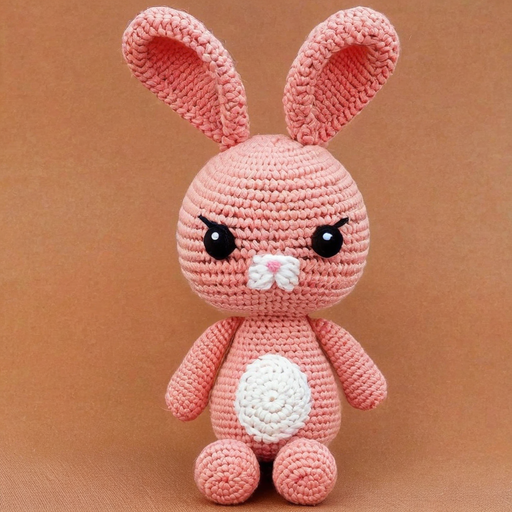}};
        \end{tikzpicture}
    \end{subfigure}
    \quad
    \begin{subfigure}[b]{0.3\textwidth}
        \begin{tikzpicture}[anchor=south west]
            \def\imagewidth{\textwidth}
            \def\imagesource{images/items/sdxl_juggernaut_xl_v8_7.png}
            \node[name=img, inner sep=0]{\includegraphics[width=\imagewidth]{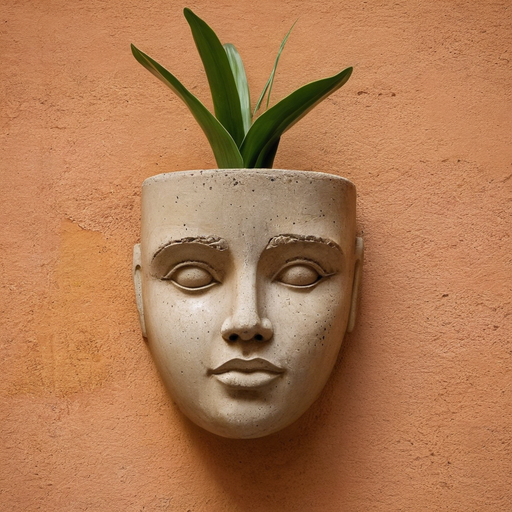}};
        \end{tikzpicture}
    \end{subfigure}

    \caption{Synthetic images of hand crafted objects in which small irregularities appear perfectly natural and thus are particularly challenging to identify as synthetic.}
    \label{fig:difficult_items}
\end{figure}